\documentclass[aps,prl,reprint,superscriptaddress]{revtex4-2}
\usepackage{amsmath}
\usepackage{graphicx}
\usepackage{float}
\usepackage{bm}
\usepackage{graphicx}   
\usepackage{multirow}  
\usepackage{booktabs}  
\usepackage[colorlinks=true, allcolors=blue]{hyperref}

\begin{document}


\title{Symmetry-Constrained Anomalous Transport in the Altermagnetic Material CuX$_2$ (X=F,Cl)}



\author{Zhengxuan Wang}
\affiliation{College of Physics, Henan Normal University,Xinxiang, Henan 453007, china}
\author{Ruqian Wu}
\email{wur@uci.edu}
\affiliation{Department of Physics and Astronomy, University of California, Irvine 92697, USA}
\author{Chunlan Ma}
\affiliation{School of Physics and Technology, Suzhou University of Science and Technology, Suzhou 215009, China}
\author{Shijing Gong}
\affiliation{Department of Physics, East China Normal University, Shanghai 200062, China}
\author{Shuaikang Zhang}
\affiliation{College of Physics, Henan Normal University,Xinxiang, Henan 453007, china}
\author{Guangtao Wang}
\affiliation{College of Physics, Henan Normal University,Xinxiang, Henan 453007, china}
\author{Tianxing Wang}
\affiliation{College of Physics, Henan Normal University,Xinxiang, Henan 453007, china}
\author{Yipeng An}
\email{ypan@htu.edu.cn}
\affiliation{College of Physics, Henan Normal University,Xinxiang, Henan 453007, china}



\date{\today}

\begin{abstract}
Recently discovered, altermagnetism represents a third class of collinear magnets. These materials exhibit zero net magnetization, similar to antiferromagnets, but display anomalous transport properties resembling those of ferromagnets. Altermagnetic materials manifest various anomalous electronic transport phenomena, including the anomalous Hall effect, anomalous Nernst effect, and anomalous thermal Hall effect. Additionally, they exhibit magneto-optical Kerr and Faraday effects, previously considered exclusive to ferromagnetic materials. These anomalous transport phenomena are constrained by symmetry, as revealed by density functional theory (DFT) calculations. However, an effective model-based approach to verify these symmetry constraints remains unavailable. In this Letter, we construct a $k\cdot p$ model for $d$-wave altermagnets CuX$_2$ (X=F,Cl) using spin space group representations and apply it to calculate the anomalous Hall effect. The symmetry-imposed transport properties predicted by the model are in agreement with the DFT results, providing a foundation for further investigation into symmetry-restricted transport phenomena in altermagnetic materials.
\end{abstract}


\maketitle
\paragraph{Introduction}
Recently, altermagnet have attracted widespread attention as a novel class of collinear magnets. Like conventional collinear antiferromagnets, altermagnet exhibit a net magnetization of zero. However, they exhibit characteristics akin to ferromagnetic materials in terms of charge and thermal transport, featuring anomalous Hall effect (AHE), anomalous Nernst effect (ANE) and anomalous thermal Hall effect (ATHE) \cite{yaoAHC}. Additionally, while systems with zero net magnetization are typically thought to be incapable of exhibiting the magneto-optical Kerr effect (MOKE) and magneto-optical Faraday (MOFE) effect, altermagnet demonstrate non-zero optical Hall conductivity and finite magneto-optical Kerr and Faraday angles \cite{yaoMOKE}.

In traditional collinear antiferromagnetic (AFM) systems, time-reversal and inversion symmetry ($\mathcal{PT}$ symmetry) guarantees spin degeneracy, as represented by $\mathcal{PT}\varphi_{k\uparrow} = \varphi_{k\downarrow}$. This symmetry spans the entire Brillouin zone, preventing spin splitting. In altermagnetic systems, however, the combined rotational and time-reversal symmetry ($\mathcal{RT}$ symmetry) replaces $\mathcal{PT}$ symmetry, but the $k$-invariant subspace of $\mathcal{RT}$ symmetry is limited to a part of the Brillouin zone. This allows spin splitting where $\mathcal{RT}$ symmetry is broken. 
Another consequence of the breaking of $\mathcal{PT}$ symmetry is the potential emergence of the anomalous Hall effect.
AHE typically requires the breaking of time-reversal symmetry, which is why it is often associated with ferromagnetic materials. However, AHE can also emerge in systems with non-zero net magnetization when non-magnetic atoms break time-reversal symmetry, as observed in materials like $\mathrm{RuO}_2$, where it was initially referred to as the crystal Hall effect \cite{yaoAHC,ruo2ahc1,ruo2ahc2}. With further study of $\mathrm{RuO}_2$, the concept of altermagnetism, a new class of collinear magnets, was introduced. Altermagnets inherently break time-reversal symmetry, making the occurrence of  AHE possible. In addition, due to the presence of $\mathcal{RT}$ and $\mathcal{P}$ symmetry operations, the components of the anomalous Hall conductivity in different directions are subject to symmetry constraints. For instance, in the case of RuO\(_2\) under the \(M_{001}\mathcal{T} \) symmetry, the AHC $\sigma_{xy}$ component is zero.

ANE \cite{ANE} and ATHE \cite{ATHE} are thermoelectric analogs of the AHE, exhibiting similar symmetry dependence \cite{AHE1,AHE2,AHE3,AHE4,AHE5,AHE6,AHE7,AHE8}. Like AHE, both ANE and ATHE arise from the breaking of time-reversal symmetry and are closely related to the system's spin-orbit coupling. Systems that exhibit AHE typically also show the ANE and ATHE, making these properties defining characteristics of altermagnets. In particular, the ATHE and ATHE, like AHE, is closely linked to the system's symmetry and demonstrates significant temperature dependence, further distinguishing altermagnetic materials from traditional AFM systems \cite{yaoAHC}. Similarly, the non-zero off-diagonal optical conductivity, as well as the magneto-optical Kerr effect and magneto-optical Faraday effect, represent unique properties of altermagnet. These magneto-optical effects provide additional signatures that set altermagnets apart from AFM materials\cite{yaoMOKE,yao3}. 

In this Letter, we investigate the CuX$_2$ (X=F,Cl) system, which crystallizes in space group No. 14. Throughout the main text, we take CuF$_2$ as a representative example, while the corresponding calculations for CuCl$_2$ are provided in the Supporting Information.
Firstly, we employed first-principles calculations to obtain the band structure of CuF\(_2\), and examined the spin splitting of its valence band on the k\(_x\)-k\(_y\) and k\(_z\)-k\(_y\) planes, which are not protected by $\mathcal{RT}$ symmetry. Subsequently, we constructed a $k\cdot p$ model based on the group representation of the spin space group of CuF\(_2\) to verify the symmetry-constrained phenomena of the anomalous Hall effect. Finally, we calculated the anomalous electrical and photoelectrical transport properties of CuF\(_2\).

The electronic structure calculations were performed using the projector augmented wave (PAW) method as implemented in the Vienna Ab Initio Simulation Package (VASP) \cite{vasp1,vasp2}. The exchange-correlation functional was treated within the generalized-gradient approximation (GGA) with the Perdew-Burke-Ernzerhof (PBE) parametrization \cite{ggapbe}.
The lattice parameters were fixed, and the atomic positions were fully relaxed until the forces on each atom were less than 0.001 $\mathrm{eV/\AA}$. The energy cutoff was set to 500 eV, with a convergence criterion of $10^{-7}$ eV. A $15 \times 11 \times 11$ k-point grid was used for the Brillouin zone sampling.
Following the self-consistent field (SCF) calculations, maximally localized Wannier functions were constructed by projecting onto $s$, $p$ and $d$ orbitals of Cu atoms as well as onto $p$ orbitals of F atoms, using a uniform k-point mesh of $17 \times 13 \times 13$ and the WANNIER90 package \cite{wannier}. Additionally, we utilized the \texttt{kdotp-generator} Python package to generate the $k\cdot p$ model.

\begin{figure}
\centering
\includegraphics[width=1\linewidth]{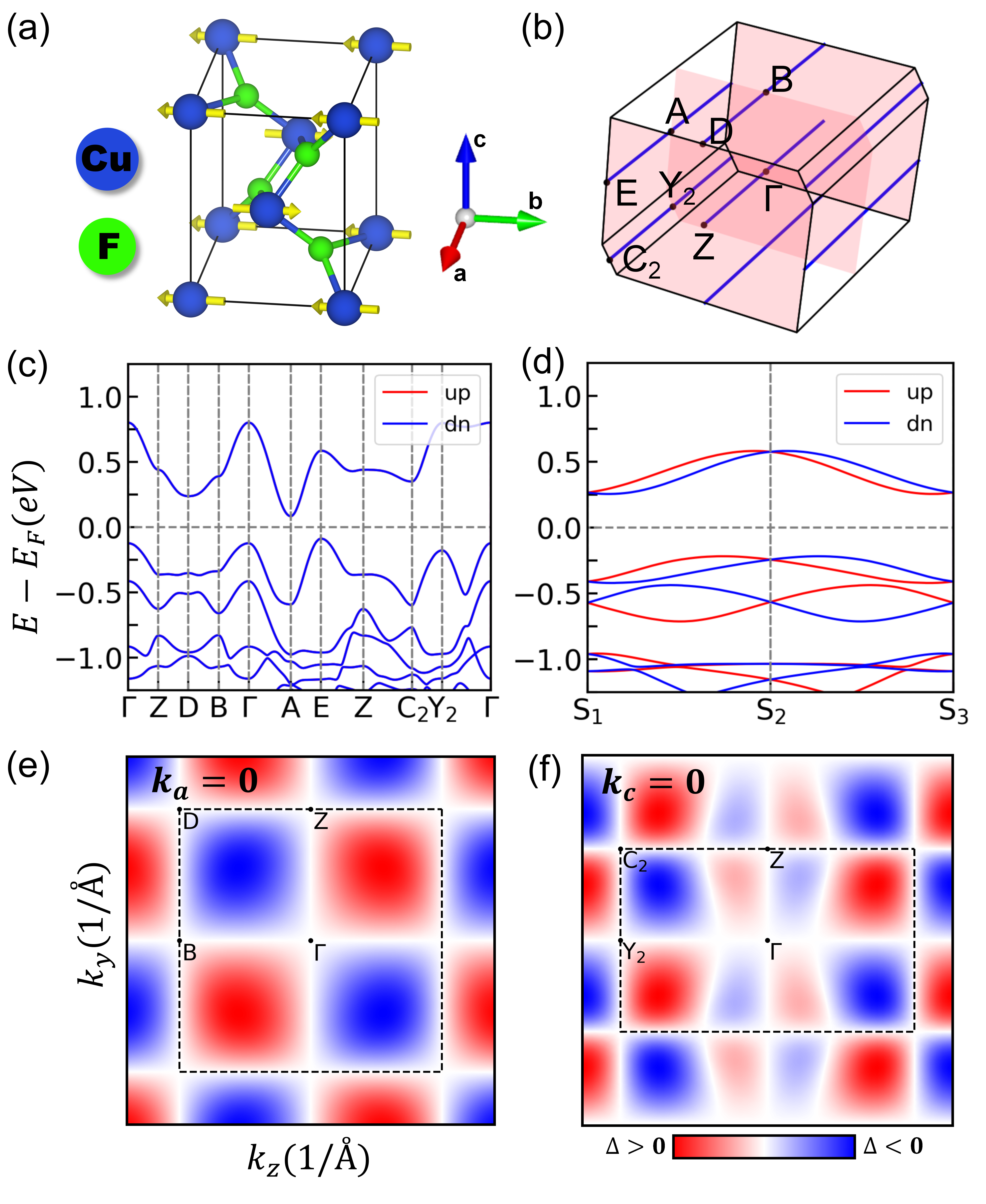}
\caption{\label{fig:band}(a), (b) The crystal structure and the Brillouin zone of $\mathrm{CuF_{2}}$ respectively. 
The red face and blue line represent the k-invariant subspace of symmetry $\{\mathcal{C}_{2y}\mathcal{T}|(0,1/2,1/2)\}$. 
(c), (d) The band structure along the path which $\bm{k}$ belong and does not belong to the invariant subspace of the symmetry operation $\{\mathcal{C}_{2y}\mathcal{T}|(0,1/2,1/2)\}$, respectively. $\mathrm{S}_1=(0,-1/2,1/2)$, $\mathrm{S}_2=(0,0,1/2)$ and $\mathrm{S}_3=(0,1/2,1/2)$.
(e) and (f) show the spin splitting in the k$_z$-k$_y$ plane and k$_x$-k$_y$ plane, respectively. The dash lines indicate the boundaries of the first Brillouin zone, and the colors correspond to $\Delta = \varepsilon_{\uparrow}-\varepsilon_{\downarrow}$.
}
\end{figure}

$\mathrm{CuF_{2}}$ has a monoclinic lattice structure with lattice constant $a=3.33$ \AA, $b=4.61$ \AA\  and $c=5.41$ \AA, the interaxial angle $\alpha=\gamma=90^\circ$ and $\beta=120^\circ$, as illustrated in  Fig \ref{fig:band}.(a).
The crystal structure of $\mathrm{CuF_{2}}$ belongs to the $P2_{1}/c$ (No.14) space group, This space group is generated by the two symmetry operations $\{\mathcal{C}_{2y}|(0,1/2,1/2)\}$ and the inversion operation, $\mathcal{P}$.
Upon reversing the magnetic moments of the two Cu atoms within the unit cell, the generating set of the system's symmetry group transforms to $\{\mathcal{C}_{2y}\mathcal{T}|(0,1/2,1/2)\}$ and $\mathcal{P}$ symmetry, where $\mathcal{T}$ represents the time-reversal operation.
Based on the space group tables \cite{spacegrouptable}, we can readily determine the $k$-invariant subspace $S$ of the $\{\mathcal{C}_{2y}\mathcal{T}|(0,1/2,1/2)\}$ symmetry operation, as illustrated in Fig \ref{fig:band}.(b). For $k$ points belonging to $S$, the relation $\{\mathcal{C}_{2y}\mathcal{T}|(0,1/2,1/2)\}\varphi_{n\bm{k}\uparrow}=\varphi_{n\bm{k}\downarrow}$ holds true, resulting in the degeneracy of spin-up and spin-down states, as shown in Fig \ref{fig:band}.(c). In contrast, for $k$ points not belonging to $S$, the spin-up and spin-down states $\varphi_{n\bm{k}\uparrow}$ and $\varphi_{n\bm{k}\downarrow}$ are not protected by the $\{\mathcal{C}_{2y}\mathcal{T}|(0,1/2,1/2)\}$ operation and are susceptible to spin splitting, as depicted in Fig \ref{fig:band}.(d).
For the $k_xk_y$ and $k_zk_y$ planes, $\mathrm{CuF_{2}}$ can be identified as a prototypical d-wave altermagnet with spin-momentum locking, as shown in Fig \ref{fig:band}.(e) and (f).

\paragraph{The model Hamilton}

\begin{table*}[ht]
    \centering
    \resizebox{\textwidth}{!}{
    \begin{tabular}{c|c|c|c|c|c|c|c|c|c|c|c|c}  
        \toprule
        & \multicolumn{3}{c|}{Operation 1} & \multicolumn{3}{c|}{Operation 2} & \multicolumn{3}{c|}{Operation 3} & \multicolumn{3}{c}{Operation 4} \\  
        \midrule
         & Rotation & Spin Rotation & Tau & Rotation & Spin Rotation & Tau & Rotation & Spin Rotation & Tau & Rotation & Spin Rotation & Tau  \\  
        \midrule
         & $\begin{pmatrix}
            1&0&0\\0&1&0\\0&0&1\\
        \end{pmatrix}$ & $\begin{pmatrix}
            1&0&0\\0&1&0\\0&0&1\\
        \end{pmatrix}$ & $\begin{pmatrix}
            0\\0\\0\\
        \end{pmatrix}$ & $\begin{pmatrix}
            -1&0&0\\0&-1&0\\0&0&-1\\
        \end{pmatrix}$ & $\begin{pmatrix}
            1&0&0\\0&1&0\\0&0&1\\
        \end{pmatrix}$ & $\begin{pmatrix}
            0\\0\\0\\
        \end{pmatrix}$ & $\begin{pmatrix}
            -1&0&0\\0&1&0\\0&0&-1\\
        \end{pmatrix}$ & $\begin{pmatrix}
            -1&0&0\\0&-1&0\\0&0&-1\\
        \end{pmatrix}$ & $\begin{pmatrix}
            0\\ \frac{1}{2} \\ \frac{1}{2} \\
        \end{pmatrix}$ & $\begin{pmatrix}
            1&0&0\\0&-1&0\\0&0&1\\
        \end{pmatrix}$ & $\begin{pmatrix}
            -1&0&0\\0&-1&0\\0&0&-1\\
        \end{pmatrix}$ & $\begin{pmatrix}
            0\\ \frac{1}{2} \\ \frac{1}{2} \\
        \end{pmatrix}$  \\  
        \midrule
        $\Gamma_1$& \multicolumn{3}{c|}{$\begin{pmatrix}
            1&0\\ 0&1 \\
        \end{pmatrix}$} & \multicolumn{3}{c|}{$\begin{pmatrix}
            1&0\\ 0&1 \\
        \end{pmatrix}$} & \multicolumn{3}{c|}{$\begin{pmatrix}
            0&\mathrm{i}\\ -\mathrm{i}&0 \\
        \end{pmatrix}$} & \multicolumn{3}{c}{$\begin{pmatrix}
            0&\mathrm{i}\\ -\mathrm{i}&0 \\
        \end{pmatrix}$} \\ 
        \midrule
        $\Gamma_2$& \multicolumn{3}{c|}{$\begin{pmatrix}
            1&0\\ 0&1 \\
        \end{pmatrix}$} & \multicolumn{3}{c|}{$\begin{pmatrix}
            -1&0\\ 0&-1 \\
        \end{pmatrix}$} & \multicolumn{3}{c|}{$\begin{pmatrix}
            0&\mathrm{i}\\ -\mathrm{i}&0 \\
        \end{pmatrix}$} & \multicolumn{3}{c}{$\begin{pmatrix}
            0&-\mathrm{i}\\ \mathrm{i}&0 \\
        \end{pmatrix}$} \\ 
        \bottomrule
    \end{tabular}
    }
    \caption{This table lists the symmetry operations of the CuF$_2$ material in the spin space group 14.1.2.3.L, along with their spin group representations. The rotation operation matrices are written in the lattice coordinate system, while the spin rotation operations are written in the Cartesian coordinate system.}
    \label{tab:ir}
\end{table*}

\begin{figure}
\centering
\includegraphics[width=1\linewidth]{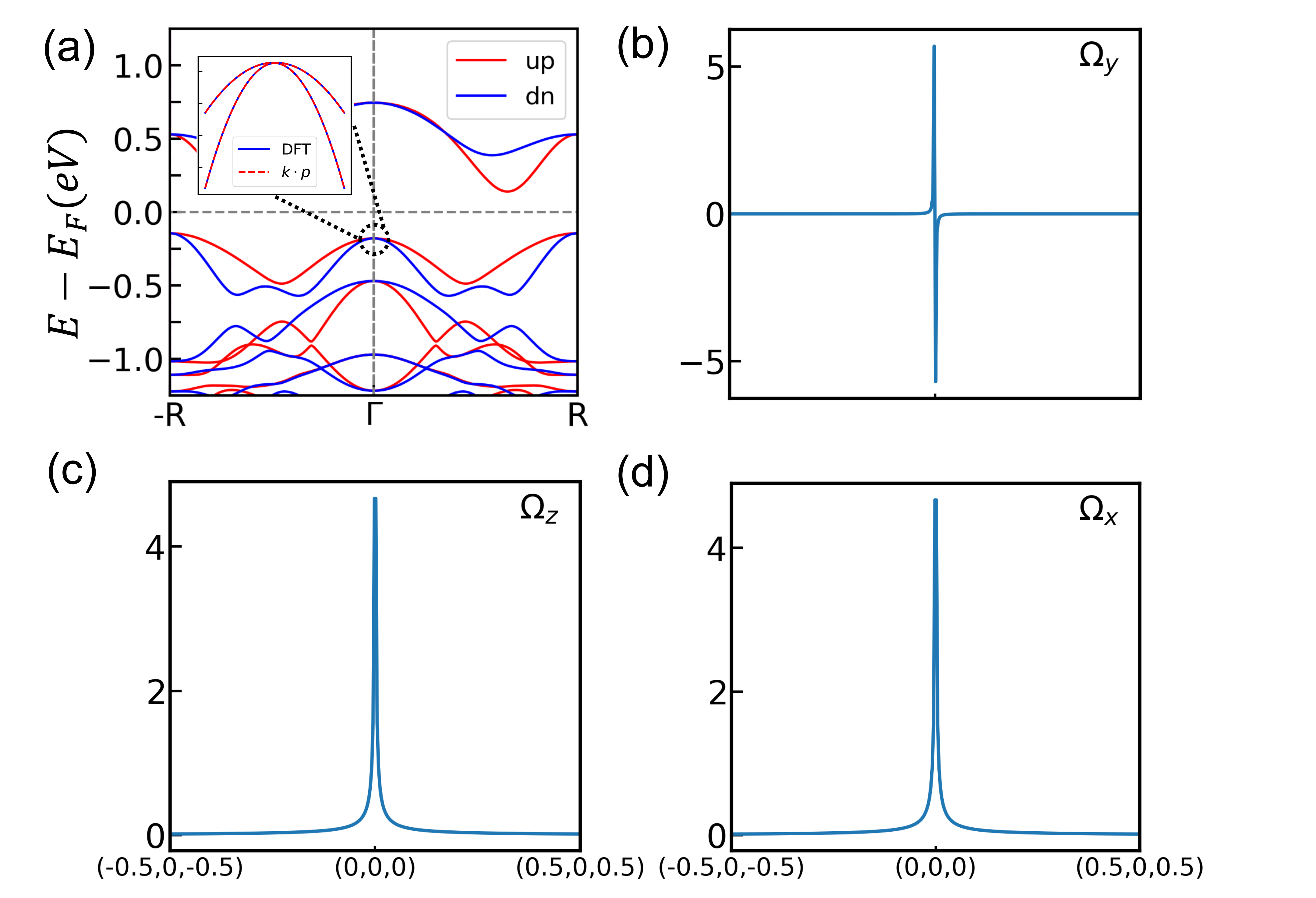}
\caption{\label{fig:model} 
(a) The band structure along the diagonal direction of the Brillouin zone passing through the $\Gamma$ point. The inset shows a comparison between the $k \cdot p$ model and DFT calculations near the $\Gamma$ point. (b), (c), and (d) are the Berry curvatures $\Omega_y$, $\Omega_x$, and $\Omega_z$ calculated on the $k_y = 0$ plane using the $k \cdot p$ model. 
}
\end{figure}

Due to the breaking of $\mathcal{PT}$ symmetry, the altermagnet cannot be fully described by their space group or magnetic space group. For instance, in the CuF$_2$ system, it belongs to 14.79 (BNS number) magnetic space group, but the irreducible representation of this magnetic space group at the $\Gamma$ point is one-dimensional, which contradicts the results obtained from our DFT calculations. Therefore, for altermagnets, we must use a spin space group with more symmetry operations to describe the symmetry constraints and construct a model based on it. The CuF$_2$ system belongs to the spin group 14.1.2.3.L, which has four symmetry operations. At the $\Gamma$ point, it has two two-dimensional irreducible representations connecting the spin-up and spin-down states, as shown in Table \ref{tab:ir}.

To obtain all the symmetries of 14.1.2.3.L, we construct the $k \cdot p$ model based on the two degenerate spin-up and spin-down states at the valence band maximum at the $\Gamma$ point, which are represented by $\Gamma_1$. The constructed $k \cdot p$ Hamiltonian is given by:

\begin{align}
H(\bm{k}) =& (c_0 + c_4k_y + c_8k_x^2 +c_{12}k_xk_z + c_{13}k_y^2 + c_{17}k_z^2)\sigma_0 \nonumber \\
&+(c_1k_x + c_5k_z + c_9k_xk_y + c_{14}k_yk_z)\sigma_1 \nonumber \\&+(c_2k_x + c_6k_z + c_{10}k_xk_y + c_{15}k_yk_z)\sigma_2 \nonumber \\
&+(c_3k_x + c_7k_z + c_{11}k_xk_y + c_{16}k_yk_z)\sigma_3
\end{align}

where $\sigma_0, \sigma_1, \sigma_2, \sigma_3$ are the Pauli matrices. We fit this Hamiltonian along the diagonal direction of the Brillouin zone with the DFT-derived energy bands, and the fitting result is shown in the small inset of Fig \ref{fig:model}.(a). For a two-band model, we can always decompose the Hamiltonian into the following form:

\begin{equation}
H(\bm{k})=\varepsilon_0(\bm{k})+\bm{h}(\bm{k})\cdot \bm{\sigma}
\end{equation}

where $\bm{h}(\bm{k})$ is the non-diagonal term of the Hamiltonian decomposed into a vector form with Pauli matrices. From this, we can obtain the Berry curvature in a simple way:

\begin{equation}
\Omega_z(\bm{k})=-\frac{1}{2h^3}\bm{h}(\bm{k})\cdot
\frac{\partial\bm{h}(\bm{k})}{\partial k_x}\times\frac{\partial\bm{h}(\bm{k})}{\partial k_y}
\end{equation}

Based on this, we calculated $\Omega_{xy}$, $\Omega_{yz}$, and $\Omega_{zx}$ on the $k_y = 0$ plane, with the results shown in Fig \ref{fig:model}(b)-(d).
It is clear that under the \( M_{010}\mathcal{T} \) symmetry operation, $\Omega_{zx}$ is always an odd function, and its integral over the Brillouin zone is always zero. This intriguing distribution of the Berry curvature arises from the \( \mathcal{T} M_{010} \) symmetry operation inherent in CuF$_2$, which directly leads to $\Omega_y$ always being an odd function along the $k_y$ direction. This can be easily derived from both the Berry curvature's Kubo formula and the aforementioned equations.
 This leads to the vanishing of the anomalous Hall conductivity component $\sigma_{zx}$.

\paragraph{Anomalous transport properties}

\begin{figure}
\centering
\includegraphics[width=1\linewidth]{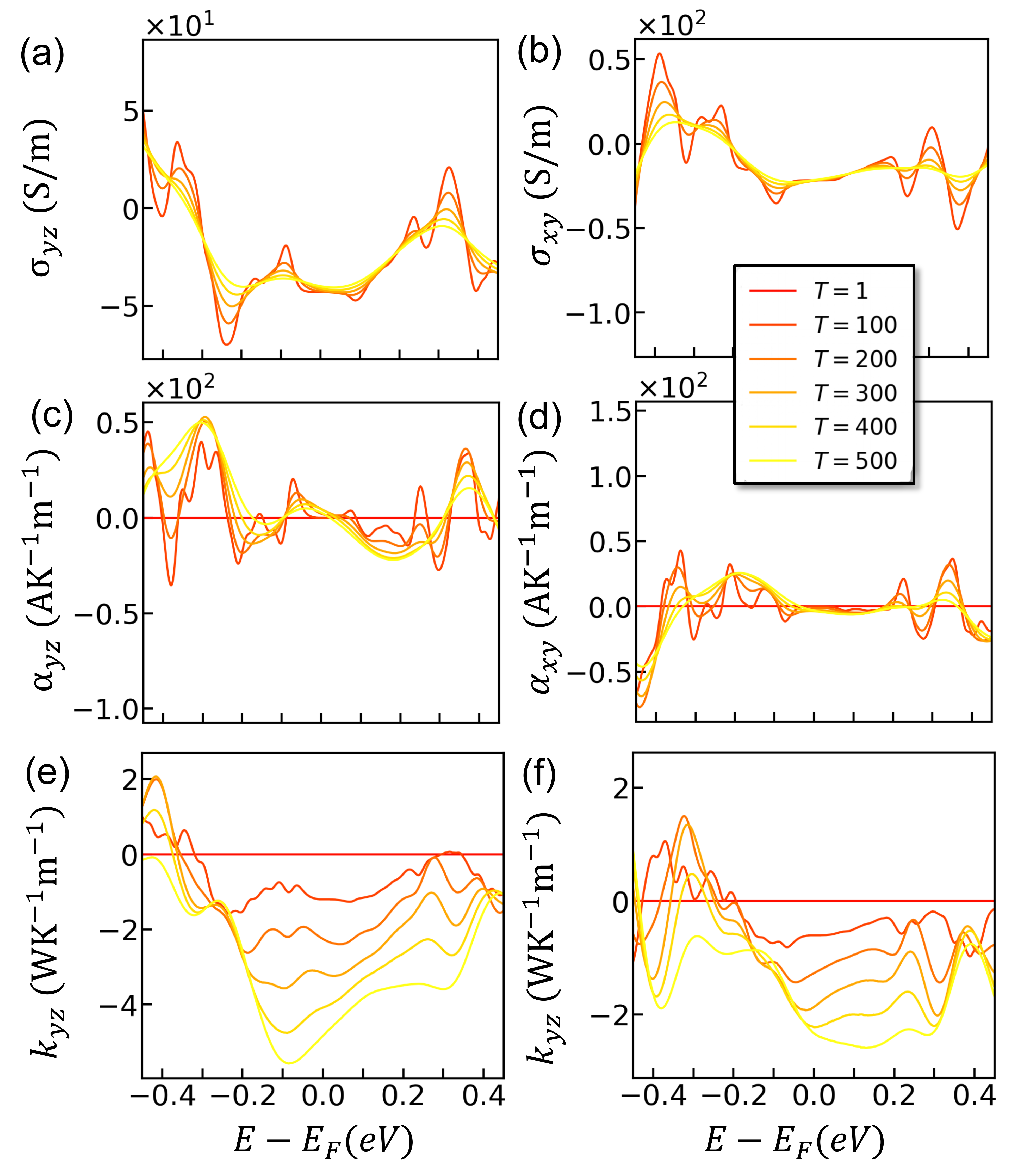}
\caption{\label{fig:ahcanc}
Temperature-dependent behavior of various transport coefficients. Panels (a) and (b) show the anomalous Hall conductivity in the $\sigma_{yz}$ and $\sigma_{xy}$ directions, respectively, as a function of the Fermi energy. Panels (c) and (d) depict the anomalous Nernst conductivity in the $\alpha_{yz}$ and $\alpha_{xy}$ directions, respectively, as a function of the Fermi energy. Panels (e) and (f) illustrate the anomalous thermal Hall conductivity in the $k_{yz}$ and $k_{xy}$ directions, respectively, as a function of the Fermi energy. In all panels, The curves in all panels represent different temperatures, with each color corresponding to a specific temperature in Kelvin. Due to the sharp peak in anomalous Hall conductivity at T = 1K, the corresponding data has been placed in the supplementary information.
}
\end{figure}

Alter-magnetic materials exhibit anomalous transport properties that distinguish them from AFM. These anomalies include non-zero anomalous Hall conductivity (AHC), anomalous Nernst conductivity (ANC), and anomalous thermal Hall conductivity (ATHC) in the absence of net magnetization. The intrinsic origin of these anomalous transport phenomena in alter-magnetic materials can be attributed to the non-zero Berry curvature integral, which arises from the energy splitting between up and down spin states in the Brillouin zone. Since these three transport properties share the same underlying symmetry, we focus on the AHC to discuss how symmetry affects these properties. The AHC, ANC and ATHC can be expressed as \cite{eqat1,eqat2,eqat3}:

\begin{eqnarray}
    \sigma_{xy}^{\mathrm{AHC}} &=& - \frac{e^2}{\hbar}\int \frac{d^3\bm{k}}{(2\pi)^3}  \Omega_{z}(\bm{k}) \label{AHC} \\
    \alpha_{xy}^{\mathrm{ANC}} &=& \int \frac{\varepsilon - \mu}{eT} \left( -\frac{\partial f}{\partial \varepsilon} \right) \sigma_{xy}^{\mathrm{AHC}} d\varepsilon \\
    k_{xy}^{\mathrm{ATHC}} &=& \int \frac{(\varepsilon - \mu)^2}{e^2T} \left( -\frac{\partial f}{\partial \varepsilon} \right) \sigma_{xy}^{\mathrm{AHC}} d\varepsilon \label{ATHC}
\end{eqnarray}
where \( \Omega_{z}(\bm{k}) \) is the momentum-resolved Berry curvature, \( \mu \) is the chemical potential, and \( f = 1/[\exp \left( (\varepsilon - \mu) / k_B T \right) + 1 ]\) is the Fermi-Dirac distribution function.

The AHC is strongly influenced by mirror symmetries, which cause certain components to vanish under specific symmetry operations. For instance, under the \( M_{001} \) symmetry, the components \( \sigma_{yz} \) and \( \sigma_{zx} \) will change sign, while \( \sigma_{xy} \) remains invariant. In this case, \( \sigma_{yz} = \sigma_{zx} = 0 \). For a combined time-reversal and mirror symmetry operation, such as \( \mathcal{T} M_{001} \), only \( \sigma_{xy} \) is affected, and in this case, \( \sigma_{xy} = 0 \). This behavior of AHC leads to the disappearance of certain components of the Hall conductivity in systems with multiple mirror symmetries, such as \( \mathrm{RuO_2} \), where the mirror symmetries are broken or restored depending on the magnetization direction\cite{yaoAHC}.

In contrast, for the \( \mathrm{CuF_2} \) system, the unique mirror symmetry is coupled with time-reversal symmetry to form the combined \( \{M_{010}\mathcal{T}|(0, 1/2, 1/2)\} \) operation. This operation always exists, regardless of the magnetization direction. As a result, the components \( \sigma_{xy} \) and \( \sigma_{yz} \) are always non-zero, while \( \sigma_{zx} \) remains zero.

The same symmetry constraints apply to the ANC and the ATHC. Since both ANC and ATHC share the same symmetry structure as AHC, their behaviors with respect to changes in magnetization direction follow a similar pattern. The changes in magnetization direction can break or restore the mirror symmetries, leading to significant modulation of the ANC and ATHC. For instance, the periodic modulation of AHC with respect to magnetization direction also manifests in ANC and ATHC, as these properties are similarly influenced by the symmetry breaking.

Based on Equations (\ref{AHC})–(\ref{ATHC}), we have calculated the temperature-dependent AHC, ANC, and ATHE, with the results shown in Figure 3. At the true Fermi level, AHC is very small and shows little variation with temperature. However, through electron or hole doping, this value can be enhanced. Within the energy range relative to the true Fermi level (from -0.5 eV to 0.5 eV), $\sigma_{yz}$ reaches a peak of -70.3 S/m at T = 100 K and -0.23 eV, while $\sigma_{xy}$ peaks at 53.2 S/m at -0.38 eV under the same temperature. For the thermoelectric coefficients, $\alpha_{yz}$ achieves a peak of 52.4 $\mathrm{AK^{-1}m^{-1}}$ at T = 300 K and -0.29 eV, and $\alpha_{xy}$ peaks at -77.2 $\mathrm{AK^{-1}m^{-1}}$ at T = 200 K and -0.44 eV. The ATHC is directly proportional to temperature, with peak values occurring at T = 500 K. Specifically, $\kappa_{yz}$ reaches a peak of -5.59 $\mathrm{WK^{-1}m^{-1}}$ at -0.08 eV, and $\kappa_{xy}$ peaks at -2.64 $\mathrm{WK^{-1}m^{-1}}$ at 0.13 eV.

\paragraph{The MOKE and MOFE}

\begin{figure}
\centering
\includegraphics[width=1\linewidth]{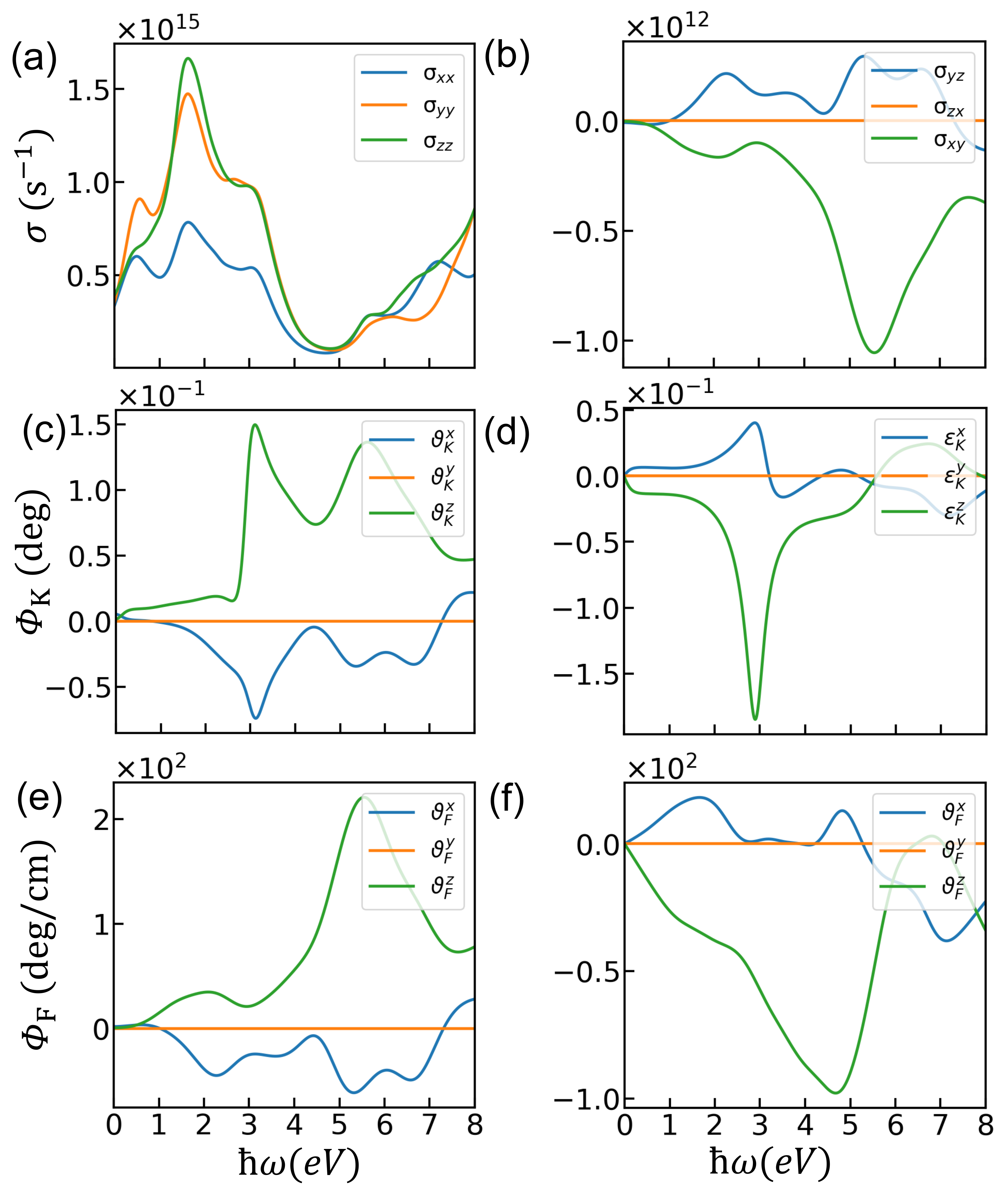}
\caption{\label{fig:moke}
Magneto-optical and optical conductivity properties. Panel (a) shows the diagonal component of the optical conductivity, while panel (b) presents the off-diagonal component. Panels (c) and (d) display the magneto-optical Kerr angle, with (c) showing the rotation angle and (d) showing the ellipticity. Panels (e) and (f) illustrate the magneto-optical Faraday angle, with (e) showing the rotation angle and (f) showing the ellipticity.
}
\end{figure}

Alternating magnetic systems with a net-zero magnetic moment exhibit optical transport properties that are more similar to those of ferromagnetic systems. 
This is evidenced by phenomena such as a non-zero optical Hall conductivity, Magneto-optical Kerr effects and Magneto-optical Faraday effects.
The optical conductivity can be given by the Kubo-Greenwood formula \cite{opteq1,opteq2},
\begin{equation}
\begin{split}
    \sigma_{xy}(\omega) =& \frac{ie^2\hbar}{N_k\Omega_c} \sum_{\bm{k}} \sum_{n,m} \frac{f_{m\bm{k}} - f_{n\bm{k}}}{\varepsilon_{m\bm{k}} - \varepsilon_{n\bm{k}}}
    \\
    &\times \frac{\langle \psi_{n\bm{k}} | v_{\alpha} | \psi_{m\bm{k}} \rangle \langle \psi_{m\bm{k}} | v_{\beta} | \psi_{n\bm{k}} \rangle }{\varepsilon_{m\bm{k}} - \varepsilon_{n\bm{k}} - ( \hbar\omega + i\eta)} ,\label{optical}
\end{split}
\end{equation}
where $\Omega_{c}$ is the cell volume, $N_k$ is the total number of $k$ points used for sampling the Brillouin zone, $\psi_{n\bm{k}}$ and $\varepsilon_{n\bm{k}}$ are respectively the Wannier function and interpolated energy at band index $n$ and momentum $\bm{k}$, $v_{\alpha,\beta}$ are velocity operators, $f_{n\bm{k}}$ is the Fermi-Dirac distribution function, $\hbar\omega$ is the photon energy, and $\eta$ is an adjustable energy smearing parameter.

For the complex Kerr and Faraday angle, we adopt a simplified expression under the assumption of a small rotation angle \cite{mokeeq1,mokeeq2,mokeeq3,mokeeq4,mokeeq5,mokeeq6,mokeeq7,mokeeq8,mokeeq9}: 
\begin{eqnarray}
    \phi^{z}_{K} = \vartheta^{z}_{K} + i\varepsilon^{z}_{K} \approx \frac{-\nu_{xyz} \sigma_{xy}}{\sigma_{0} \sqrt{1 + i(4\pi/\omega)\sigma_{0}}} ,  \\
    \phi^{z}_{F} = \vartheta^{z}_{F} + i\varepsilon^{z}_{K} \approx \frac{-\nu_{xyz} \sigma_{xy}}{ \sqrt{1 + i(4\pi/\omega)\sigma_{0}}} \frac{2\pi}{c} , 
\end{eqnarray}
where $\sigma_0 = (\sigma_{xx} +\sigma_{yy})/2$.
Detailed calculation steps can be found in the supplementary information.


Based on Equation (\ref{optical}), when the system exhibits mirror time-reversal (\( \{M_{010}\mathcal{T}|(0, 1/2, 1/2)\} \)) symmetry, the off-diagonal components of the optical conductivity tensor are constrained by this symmetry. In this case, the off-diagonal element $\sigma_{zx}$ changes sign under \( \{M_{010}\mathcal{T}|(0, 1/2, 1/2)\} \) symmetry, while $\sigma_{xy}$ and $\sigma_{yz}$ remain invariant, leading to the disappearance of $\sigma_{zx}$, as shown in Figure \ref{fig:moke}. Since both the Kerr and Faraday angles are derived from off-diagonal terms, their spectra exhibit similar behavior, i.e., $\phi_{Kky} = 0$ and $\phi_{Fky} = 0$, as shown in panels (c)-(f) of Figure \ref{fig:moke}.(b). 
According to our calculations, the peak value of the Kerr angle in the $z$ direction is approximately 0.15 deg at 3.09 eV, while in the $x$ direction, the peak value is around -0.07 deg at 3.18 eV. The peak value of the Faraday angle in the $z$ direction is approximately 221 deg/cm at 5.54 eV, while in the $x$ direction, the peak value is around 62 deg/cm at 5.31 eV.

In summary, altermagnets exhibit transport properties distinct from those of traditional collinear magnetic materials, and these transport properties are strictly constrained by symmetry. Moreover, neither the space group nor the magnetic space group can fully describe the symmetry constraints of altermagnets. We must use a spin space group with more comprehensive symmetry operations to construct a model of altermagnets with complete symmetry. This provides a method for further studying the symmetry-constrained transport properties of altermagnets.

We acknowledge funding from the National Natural Science Foundation of China (Grant Nos. 12274117, 62274066, and 62275074), the Program for Innovative Research Team (in Science and Technology) in University of Henan Province (Grant No. 24IRTSTHN025), the Natural Science Foundation of Henan (Grant No. 242300421214), and the HPCC of HNU. Work at UCI was supported by the U.S. DOE, Office of Science (Grant No. DE-FG02-05ER46237).


\begin{thebibliography}{34}%
\makeatletter
\providecommand \@ifxundefined [1]{%
 \@ifx{#1\undefined}
}%
\providecommand \@ifnum [1]{%
 \ifnum #1\expandafter \@firstoftwo
 \else \expandafter \@secondoftwo
 \fi
}%
\providecommand \@ifx [1]{%
 \ifx #1\expandafter \@firstoftwo
 \else \expandafter \@secondoftwo
 \fi
}%
\providecommand \natexlab [1]{#1}%
\providecommand \enquote  [1]{``#1''}%
\providecommand \bibnamefont  [1]{#1}%
\providecommand \bibfnamefont [1]{#1}%
\providecommand \citenamefont [1]{#1}%
\providecommand \href@noop [0]{\@secondoftwo}%
\providecommand \href [0]{\begingroup \@sanitize@url \@href}%
\providecommand \@href[1]{\@@startlink{#1}\@@href}%
\providecommand \@@href[1]{\endgroup#1\@@endlink}%
\providecommand \@sanitize@url [0]{\catcode `\\12\catcode `\$12\catcode
  `\&12\catcode `\#12\catcode `\^12\catcode `\_12\catcode `\%12\relax}%
\providecommand \@@startlink[1]{}%
\providecommand \@@endlink[0]{}%
\providecommand \url  [0]{\begingroup\@sanitize@url \@url }%
\providecommand \@url [1]{\endgroup\@href {#1}{\urlprefix }}%
\providecommand \urlprefix  [0]{URL }%
\providecommand \Eprint [0]{\href }%
\providecommand \doibase [0]{https://doi.org/}%
\providecommand \selectlanguage [0]{\@gobble}%
\providecommand \bibinfo  [0]{\@secondoftwo}%
\providecommand \bibfield  [0]{\@secondoftwo}%
\providecommand \translation [1]{[#1]}%
\providecommand \BibitemOpen [0]{}%
\providecommand \bibitemStop [0]{}%
\providecommand \bibitemNoStop [0]{.\EOS\space}%
\providecommand \EOS [0]{\spacefactor3000\relax}%
\providecommand \BibitemShut  [1]{\csname bibitem#1\endcsname}%
\let\auto@bib@innerbib\@empty
\bibitem [{\citenamefont {Zhou}\ \emph {et~al.}(2024)\citenamefont {Zhou},
  \citenamefont {Feng}, \citenamefont {Zhang}, \citenamefont
  {\ifmmode~\check{S}\else \v{S}\fi{}mejkal}, \citenamefont {Sinova},
  \citenamefont {Mokrousov},\ and\ \citenamefont {Yao}}]{yaoAHC}%
  \BibitemOpen
  \bibfield  {author} {\bibinfo {author} {\bibfnamefont {X.}~\bibnamefont
  {Zhou}}, \bibinfo {author} {\bibfnamefont {W.}~\bibnamefont {Feng}}, \bibinfo
  {author} {\bibfnamefont {R.-W.}\ \bibnamefont {Zhang}}, \bibinfo {author}
  {\bibfnamefont {L.}~\bibnamefont {\ifmmode~\check{S}\else \v{S}\fi{}mejkal}},
  \bibinfo {author} {\bibfnamefont {J.}~\bibnamefont {Sinova}}, \bibinfo
  {author} {\bibfnamefont {Y.}~\bibnamefont {Mokrousov}},\ and\ \bibinfo
  {author} {\bibfnamefont {Y.}~\bibnamefont {Yao}},\ }\bibfield  {title}
  {\bibinfo {title} {Crystal thermal transport in altermagnetic
  ${\mathrm{ruo}}_{2}$},\ }\href
  {https://doi.org/10.1103/PhysRevLett.132.056701} {\bibfield  {journal}
  {\bibinfo  {journal} {Phys. Rev. Lett.}\ }\textbf {\bibinfo {volume} {132}},\
  \bibinfo {pages} {056701} (\bibinfo {year} {2024})}\BibitemShut {NoStop}%
\bibitem [{\citenamefont {Zhou}\ \emph {et~al.}(2021)\citenamefont {Zhou},
  \citenamefont {Feng}, \citenamefont {Yang}, \citenamefont {Guo},\ and\
  \citenamefont {Yao}}]{yaoMOKE}%
  \BibitemOpen
  \bibfield  {author} {\bibinfo {author} {\bibfnamefont {X.}~\bibnamefont
  {Zhou}}, \bibinfo {author} {\bibfnamefont {W.}~\bibnamefont {Feng}}, \bibinfo
  {author} {\bibfnamefont {X.}~\bibnamefont {Yang}}, \bibinfo {author}
  {\bibfnamefont {G.-Y.}\ \bibnamefont {Guo}},\ and\ \bibinfo {author}
  {\bibfnamefont {Y.}~\bibnamefont {Yao}},\ }\bibfield  {title} {\bibinfo
  {title} {Crystal chirality magneto-optical effects in collinear
  antiferromagnets},\ }\href {https://doi.org/10.1103/PhysRevB.104.024401}
  {\bibfield  {journal} {\bibinfo  {journal} {Phys. Rev. B}\ }\textbf {\bibinfo
  {volume} {104}},\ \bibinfo {pages} {024401} (\bibinfo {year}
  {2021})}\BibitemShut {NoStop}%
\bibitem [{\citenamefont {Šmejkal}\ \emph {et~al.}(2020)\citenamefont
  {Šmejkal}, \citenamefont {González-Hernández}, \citenamefont {Jungwirth},\
  and\ \citenamefont {Sinova}}]{ruo2ahc1}%
  \BibitemOpen
  \bibfield  {author} {\bibinfo {author} {\bibfnamefont {L.}~\bibnamefont
  {Šmejkal}}, \bibinfo {author} {\bibfnamefont {R.}~\bibnamefont
  {González-Hernández}}, \bibinfo {author} {\bibfnamefont {T.}~\bibnamefont
  {Jungwirth}},\ and\ \bibinfo {author} {\bibfnamefont {J.}~\bibnamefont
  {Sinova}},\ }\bibfield  {title} {\bibinfo {title} {Crystal time-reversal
  symmetry breaking and spontaneous hall effect in collinear
  antiferromagnets},\ }\href {https://doi.org/10.1126/sciadv.aaz8809}
  {\bibfield  {journal} {\bibinfo  {journal} {Science Advances}\ }\textbf
  {\bibinfo {volume} {6}},\ \bibinfo {pages} {eaaz8809} (\bibinfo {year}
  {2020})}\BibitemShut {NoStop}%
\bibitem [{\citenamefont {Feng}\ \emph {et~al.}(2022)\citenamefont {Feng},
  \citenamefont {Zhou}, \citenamefont {{\v{S}}mejkal}, \citenamefont {Wu},
  \citenamefont {Zhu}, \citenamefont {Guo}, \citenamefont
  {Gonz{\'a}lez-Hern{\'a}ndez}, \citenamefont {Wang}, \citenamefont {Yan},
  \citenamefont {Qin} \emph {et~al.}}]{ruo2ahc2}%
  \BibitemOpen
  \bibfield  {author} {\bibinfo {author} {\bibfnamefont {Z.}~\bibnamefont
  {Feng}}, \bibinfo {author} {\bibfnamefont {X.}~\bibnamefont {Zhou}}, \bibinfo
  {author} {\bibfnamefont {L.}~\bibnamefont {{\v{S}}mejkal}}, \bibinfo {author}
  {\bibfnamefont {L.}~\bibnamefont {Wu}}, \bibinfo {author} {\bibfnamefont
  {Z.}~\bibnamefont {Zhu}}, \bibinfo {author} {\bibfnamefont {H.}~\bibnamefont
  {Guo}}, \bibinfo {author} {\bibfnamefont {R.}~\bibnamefont
  {Gonz{\'a}lez-Hern{\'a}ndez}}, \bibinfo {author} {\bibfnamefont
  {X.}~\bibnamefont {Wang}}, \bibinfo {author} {\bibfnamefont {H.}~\bibnamefont
  {Yan}}, \bibinfo {author} {\bibfnamefont {P.}~\bibnamefont {Qin}}, \emph
  {et~al.},\ }\bibfield  {title} {\bibinfo {title} {An anomalous hall effect in
  altermagnetic ruthenium dioxide},\ }\href
  {https://doi.org/10.1038/s41928-022-00866-z} {\bibfield  {journal} {\bibinfo
  {journal} {Nature Electronics}\ }\textbf {\bibinfo {volume} {5}},\ \bibinfo
  {pages} {735} (\bibinfo {year} {2022})}\BibitemShut {NoStop}%
\bibitem [{\citenamefont {Xiao}\ \emph {et~al.}(2006)\citenamefont {Xiao},
  \citenamefont {Yao}, \citenamefont {Fang},\ and\ \citenamefont {Niu}}]{ANE}%
  \BibitemOpen
  \bibfield  {author} {\bibinfo {author} {\bibfnamefont {D.}~\bibnamefont
  {Xiao}}, \bibinfo {author} {\bibfnamefont {Y.}~\bibnamefont {Yao}}, \bibinfo
  {author} {\bibfnamefont {Z.}~\bibnamefont {Fang}},\ and\ \bibinfo {author}
  {\bibfnamefont {Q.}~\bibnamefont {Niu}},\ }\bibfield  {title} {\bibinfo
  {title} {Berry-phase effect in anomalous thermoelectric transport},\ }\href
  {https://doi.org/10.1103/PhysRevLett.97.026603} {\bibfield  {journal}
  {\bibinfo  {journal} {Phys. Rev. Lett.}\ }\textbf {\bibinfo {volume} {97}},\
  \bibinfo {pages} {026603} (\bibinfo {year} {2006})}\BibitemShut {NoStop}%
\bibitem [{\citenamefont {Qin}\ \emph {et~al.}(2011)\citenamefont {Qin},
  \citenamefont {Niu},\ and\ \citenamefont {Shi}}]{ATHE}%
  \BibitemOpen
  \bibfield  {author} {\bibinfo {author} {\bibfnamefont {T.}~\bibnamefont
  {Qin}}, \bibinfo {author} {\bibfnamefont {Q.}~\bibnamefont {Niu}},\ and\
  \bibinfo {author} {\bibfnamefont {J.}~\bibnamefont {Shi}},\ }\bibfield
  {title} {\bibinfo {title} {Energy magnetization and the thermal hall
  effect},\ }\href {https://doi.org/10.1103/PhysRevLett.107.236601} {\bibfield
  {journal} {\bibinfo  {journal} {Phys. Rev. Lett.}\ }\textbf {\bibinfo
  {volume} {107}},\ \bibinfo {pages} {236601} (\bibinfo {year}
  {2011})}\BibitemShut {NoStop}%
\bibitem [{\citenamefont {Onose}\ \emph {et~al.}(2008)\citenamefont {Onose},
  \citenamefont {Shiomi},\ and\ \citenamefont {Tokura}}]{AHE1}%
  \BibitemOpen
  \bibfield  {author} {\bibinfo {author} {\bibfnamefont {Y.}~\bibnamefont
  {Onose}}, \bibinfo {author} {\bibfnamefont {Y.}~\bibnamefont {Shiomi}},\ and\
  \bibinfo {author} {\bibfnamefont {Y.}~\bibnamefont {Tokura}},\ }\bibfield
  {title} {\bibinfo {title} {Lorenz number determination of the dissipationless
  nature of the anomalous hall effect in itinerant ferromagnets},\ }\href
  {https://doi.org/10.1103/PhysRevLett.100.016601} {\bibfield  {journal}
  {\bibinfo  {journal} {Phys. Rev. Lett.}\ }\textbf {\bibinfo {volume} {100}},\
  \bibinfo {pages} {016601} (\bibinfo {year} {2008})}\BibitemShut {NoStop}%
\bibitem [{\citenamefont {Shiomi}\ \emph {et~al.}(2009)\citenamefont {Shiomi},
  \citenamefont {Onose},\ and\ \citenamefont {Tokura}}]{AHE2}%
  \BibitemOpen
  \bibfield  {author} {\bibinfo {author} {\bibfnamefont {Y.}~\bibnamefont
  {Shiomi}}, \bibinfo {author} {\bibfnamefont {Y.}~\bibnamefont {Onose}},\ and\
  \bibinfo {author} {\bibfnamefont {Y.}~\bibnamefont {Tokura}},\ }\bibfield
  {title} {\bibinfo {title} {Extrinsic anomalous hall effect in charge and heat
  transport in pure iron, ${\text{fe}}_{0.997}{\text{si}}_{0.003}$, and
  ${\text{fe}}_{0.97}{\text{co}}_{0.03}$},\ }\href
  {https://doi.org/10.1103/PhysRevB.79.100404} {\bibfield  {journal} {\bibinfo
  {journal} {Phys. Rev. B}\ }\textbf {\bibinfo {volume} {79}},\ \bibinfo
  {pages} {100404} (\bibinfo {year} {2009})}\BibitemShut {NoStop}%
\bibitem [{\citenamefont {Shiomi}\ \emph {et~al.}(2010)\citenamefont {Shiomi},
  \citenamefont {Onose},\ and\ \citenamefont {Tokura}}]{AHE3}%
  \BibitemOpen
  \bibfield  {author} {\bibinfo {author} {\bibfnamefont {Y.}~\bibnamefont
  {Shiomi}}, \bibinfo {author} {\bibfnamefont {Y.}~\bibnamefont {Onose}},\ and\
  \bibinfo {author} {\bibfnamefont {Y.}~\bibnamefont {Tokura}},\ }\bibfield
  {title} {\bibinfo {title} {Effect of scattering on intrinsic anomalous hall
  effect investigated by lorenz ratio},\ }\href
  {https://doi.org/10.1103/PhysRevB.81.054414} {\bibfield  {journal} {\bibinfo
  {journal} {Phys. Rev. B}\ }\textbf {\bibinfo {volume} {81}},\ \bibinfo
  {pages} {054414} (\bibinfo {year} {2010})}\BibitemShut {NoStop}%
\bibitem [{\citenamefont {Smith}(1911)}]{AHE4}%
  \BibitemOpen
  \bibfield  {author} {\bibinfo {author} {\bibfnamefont {A.~W.}\ \bibnamefont
  {Smith}},\ }\bibfield  {title} {\bibinfo {title} {The transverse
  thermomagnetic effect in nickel and cobalt},\ }\href
  {https://doi.org/10.1103/PhysRevSeriesI.33.295} {\bibfield  {journal}
  {\bibinfo  {journal} {Phys. Rev. (Series I)}\ }\textbf {\bibinfo {volume}
  {33}},\ \bibinfo {pages} {295} (\bibinfo {year} {1911})}\BibitemShut
  {NoStop}%
\bibitem [{\citenamefont {Smith}(1921)}]{AHE5}%
  \BibitemOpen
  \bibfield  {author} {\bibinfo {author} {\bibfnamefont {A.~W.}\ \bibnamefont
  {Smith}},\ }\bibfield  {title} {\bibinfo {title} {The hall effect and the
  nernst effect in magnetic alloys},\ }\href
  {https://doi.org/10.1103/PhysRev.17.23} {\bibfield  {journal} {\bibinfo
  {journal} {Phys. Rev.}\ }\textbf {\bibinfo {volume} {17}},\ \bibinfo {pages}
  {23} (\bibinfo {year} {1921})}\BibitemShut {NoStop}%
\bibitem [{\citenamefont {Lee}\ \emph {et~al.}(2004)\citenamefont {Lee},
  \citenamefont {Watauchi}, \citenamefont {Miller}, \citenamefont {Cava},\ and\
  \citenamefont {Ong}}]{AHE6}%
  \BibitemOpen
  \bibfield  {author} {\bibinfo {author} {\bibfnamefont {W.-L.}\ \bibnamefont
  {Lee}}, \bibinfo {author} {\bibfnamefont {S.}~\bibnamefont {Watauchi}},
  \bibinfo {author} {\bibfnamefont {V.~L.}\ \bibnamefont {Miller}}, \bibinfo
  {author} {\bibfnamefont {R.~J.}\ \bibnamefont {Cava}},\ and\ \bibinfo
  {author} {\bibfnamefont {N.~P.}\ \bibnamefont {Ong}},\ }\bibfield  {title}
  {\bibinfo {title} {Anomalous hall heat current and nernst effect in the
  ${\mathrm{c}\mathrm{u}\mathrm{c}\mathrm{r}}_{2}{\mathrm{s}\mathrm{e}}_{4\ensuremath{-}x}{\mathrm{b}\mathrm{r}}_{x}$
  ferromagnet},\ }\href {https://doi.org/10.1103/PhysRevLett.93.226601}
  {\bibfield  {journal} {\bibinfo  {journal} {Phys. Rev. Lett.}\ }\textbf
  {\bibinfo {volume} {93}},\ \bibinfo {pages} {226601} (\bibinfo {year}
  {2004})}\BibitemShut {NoStop}%
\bibitem [{\citenamefont {Miyasato}\ \emph {et~al.}(2007)\citenamefont
  {Miyasato}, \citenamefont {Abe}, \citenamefont {Fujii}, \citenamefont
  {Asamitsu}, \citenamefont {Onoda}, \citenamefont {Onose}, \citenamefont
  {Nagaosa},\ and\ \citenamefont {Tokura}}]{AHE7}%
  \BibitemOpen
  \bibfield  {author} {\bibinfo {author} {\bibfnamefont {T.}~\bibnamefont
  {Miyasato}}, \bibinfo {author} {\bibfnamefont {N.}~\bibnamefont {Abe}},
  \bibinfo {author} {\bibfnamefont {T.}~\bibnamefont {Fujii}}, \bibinfo
  {author} {\bibfnamefont {A.}~\bibnamefont {Asamitsu}}, \bibinfo {author}
  {\bibfnamefont {S.}~\bibnamefont {Onoda}}, \bibinfo {author} {\bibfnamefont
  {Y.}~\bibnamefont {Onose}}, \bibinfo {author} {\bibfnamefont
  {N.}~\bibnamefont {Nagaosa}},\ and\ \bibinfo {author} {\bibfnamefont
  {Y.}~\bibnamefont {Tokura}},\ }\bibfield  {title} {\bibinfo {title}
  {Crossover behavior of the anomalous hall effect and anomalous nernst effect
  in itinerant ferromagnets},\ }\href
  {https://doi.org/10.1103/PhysRevLett.99.086602} {\bibfield  {journal}
  {\bibinfo  {journal} {Phys. Rev. Lett.}\ }\textbf {\bibinfo {volume} {99}},\
  \bibinfo {pages} {086602} (\bibinfo {year} {2007})}\BibitemShut {NoStop}%
\bibitem [{\citenamefont {Onoda}\ \emph {et~al.}(2008)\citenamefont {Onoda},
  \citenamefont {Sugimoto},\ and\ \citenamefont {Nagaosa}}]{AHE8}%
  \BibitemOpen
  \bibfield  {author} {\bibinfo {author} {\bibfnamefont {S.}~\bibnamefont
  {Onoda}}, \bibinfo {author} {\bibfnamefont {N.}~\bibnamefont {Sugimoto}},\
  and\ \bibinfo {author} {\bibfnamefont {N.}~\bibnamefont {Nagaosa}},\
  }\bibfield  {title} {\bibinfo {title} {Quantum transport theory of anomalous
  electric, thermoelectric, and thermal hall effects in ferromagnets},\ }\href
  {https://doi.org/10.1103/PhysRevB.77.165103} {\bibfield  {journal} {\bibinfo
  {journal} {Phys. Rev. B}\ }\textbf {\bibinfo {volume} {77}},\ \bibinfo
  {pages} {165103} (\bibinfo {year} {2008})}\BibitemShut {NoStop}%
\bibitem [{\citenamefont {Bai}\ \emph {et~al.}(2024)\citenamefont {Bai},
  \citenamefont {Feng}, \citenamefont {Liu}, \citenamefont {Šmejkal},
  \citenamefont {Mokrousov},\ and\ \citenamefont {Yao}}]{yao3}%
  \BibitemOpen
  \bibfield  {author} {\bibinfo {author} {\bibfnamefont {L.}~\bibnamefont
  {Bai}}, \bibinfo {author} {\bibfnamefont {W.}~\bibnamefont {Feng}}, \bibinfo
  {author} {\bibfnamefont {S.}~\bibnamefont {Liu}}, \bibinfo {author}
  {\bibfnamefont {L.}~\bibnamefont {Šmejkal}}, \bibinfo {author}
  {\bibfnamefont {Y.}~\bibnamefont {Mokrousov}},\ and\ \bibinfo {author}
  {\bibfnamefont {Y.}~\bibnamefont {Yao}},\ }\bibfield  {title} {\bibinfo
  {title} {Altermagnetism: Exploring new frontiers in magnetism and
  spintronics},\ }\href
  {https://doi.org/https://doi.org/10.1002/adfm.202409327} {\bibfield
  {journal} {\bibinfo  {journal} {Advanced Functional Materials}\ }\textbf
  {\bibinfo {volume} {34}},\ \bibinfo {pages} {2409327} (\bibinfo {year}
  {2024})}\BibitemShut {NoStop}%
\bibitem [{\citenamefont {Kresse}\ and\ \citenamefont
  {Furthmüller}(1996)}]{vasp1}%
  \BibitemOpen
  \bibfield  {author} {\bibinfo {author} {\bibfnamefont {G.}~\bibnamefont
  {Kresse}}\ and\ \bibinfo {author} {\bibfnamefont {J.}~\bibnamefont
  {Furthmüller}},\ }\bibfield  {title} {\bibinfo {title} {Efficiency of
  ab-initio total energy calculations for metals and semiconductors using a
  plane-wave basis set},\ }\href
  {https://doi.org/https://doi.org/10.1016/0927-0256(96)00008-0} {\bibfield
  {journal} {\bibinfo  {journal} {Computational Materials Science}\ }\textbf
  {\bibinfo {volume} {6}},\ \bibinfo {pages} {15} (\bibinfo {year}
  {1996})}\BibitemShut {NoStop}%
\bibitem [{\citenamefont {Bl\"ochl}(1994)}]{vasp2}%
  \BibitemOpen
  \bibfield  {author} {\bibinfo {author} {\bibfnamefont {P.~E.}\ \bibnamefont
  {Bl\"ochl}},\ }\bibfield  {title} {\bibinfo {title} {Projector augmented-wave
  method},\ }\href {https://doi.org/10.1103/PhysRevB.50.17953} {\bibfield
  {journal} {\bibinfo  {journal} {Phys. Rev. B}\ }\textbf {\bibinfo {volume}
  {50}},\ \bibinfo {pages} {17953} (\bibinfo {year} {1994})}\BibitemShut
  {NoStop}%
\bibitem [{\citenamefont {Perdew}\ \emph {et~al.}(1996)\citenamefont {Perdew},
  \citenamefont {Burke},\ and\ \citenamefont {Ernzerhof}}]{ggapbe}%
  \BibitemOpen
  \bibfield  {author} {\bibinfo {author} {\bibfnamefont {J.~P.}\ \bibnamefont
  {Perdew}}, \bibinfo {author} {\bibfnamefont {K.}~\bibnamefont {Burke}},\ and\
  \bibinfo {author} {\bibfnamefont {M.}~\bibnamefont {Ernzerhof}},\ }\bibfield
  {title} {\bibinfo {title} {Generalized gradient approximation made simple},\
  }\href {https://doi.org/10.1103/PhysRevLett.77.3865} {\bibfield  {journal}
  {\bibinfo  {journal} {Phys. Rev. Lett.}\ }\textbf {\bibinfo {volume} {77}},\
  \bibinfo {pages} {3865} (\bibinfo {year} {1996})}\BibitemShut {NoStop}%
\bibitem [{\citenamefont {Mostofi}\ \emph
  {et~al.}(2008{\natexlab{a}})\citenamefont {Mostofi}, \citenamefont {Yates},
  \citenamefont {Lee}, \citenamefont {Souza}, \citenamefont {Vanderbilt},\ and\
  \citenamefont {Marzari}}]{wannier}%
  \BibitemOpen
  \bibfield  {author} {\bibinfo {author} {\bibfnamefont {A.~A.}\ \bibnamefont
  {Mostofi}}, \bibinfo {author} {\bibfnamefont {J.~R.}\ \bibnamefont {Yates}},
  \bibinfo {author} {\bibfnamefont {Y.-S.}\ \bibnamefont {Lee}}, \bibinfo
  {author} {\bibfnamefont {I.}~\bibnamefont {Souza}}, \bibinfo {author}
  {\bibfnamefont {D.}~\bibnamefont {Vanderbilt}},\ and\ \bibinfo {author}
  {\bibfnamefont {N.}~\bibnamefont {Marzari}},\ }\bibfield  {title} {\bibinfo
  {title} {wannier90: A tool for obtaining maximally-localised wannier
  functions},\ }\href
  {https://doi.org/https://doi.org/10.1016/j.cpc.2007.11.016} {\bibfield
  {journal} {\bibinfo  {journal} {Computer Physics Communications}\ }\textbf
  {\bibinfo {volume} {178}},\ \bibinfo {pages} {685} (\bibinfo {year}
  {2008}{\natexlab{a}})}\BibitemShut {NoStop}%
\bibitem [{\citenamefont {Aroyo}\ \emph {et~al.}(2006)\citenamefont {Aroyo},
  \citenamefont {Kirov}, \citenamefont {Capillas}, \citenamefont {Perez-Mato},\
  and\ \citenamefont {Wondratschek}}]{spacegrouptable}%
  \BibitemOpen
  \bibfield  {author} {\bibinfo {author} {\bibfnamefont {M.~I.}\ \bibnamefont
  {Aroyo}}, \bibinfo {author} {\bibfnamefont {A.}~\bibnamefont {Kirov}},
  \bibinfo {author} {\bibfnamefont {C.}~\bibnamefont {Capillas}}, \bibinfo
  {author} {\bibfnamefont {J.~M.}\ \bibnamefont {Perez-Mato}},\ and\ \bibinfo
  {author} {\bibfnamefont {H.}~\bibnamefont {Wondratschek}},\ }\bibfield
  {title} {\bibinfo {title} {{Bilbao Crystallographic Server. II.
  Representations of crystallographic point groups and space groups}},\ }\href
  {https://doi.org/10.1107/S0108767305040286} {\bibfield  {journal} {\bibinfo
  {journal} {Acta Crystallographica Section A}\ }\textbf {\bibinfo {volume}
  {62}},\ \bibinfo {pages} {115} (\bibinfo {year} {2006})}\BibitemShut
  {NoStop}%
\bibitem [{\citenamefont {Ehrenreich}\ and\ \citenamefont
  {Spaepen}(2001)}]{eqat1}%
  \BibitemOpen
  \bibfield  {author} {\bibinfo {author} {\bibfnamefont {H.}~\bibnamefont
  {Ehrenreich}}\ and\ \bibinfo {author} {\bibfnamefont {F.}~\bibnamefont
  {Spaepen}},\ }\href@noop {} {\emph {\bibinfo {title} {Solid state
  physics}}},\ Vol.~\bibinfo {volume} {56}\ (\bibinfo  {publisher} {Academic
  Press},\ \bibinfo {year} {2001})\BibitemShut {NoStop}%
\bibitem [{\citenamefont {Behnia}(2015)}]{eqat2}%
  \BibitemOpen
  \bibfield  {author} {\bibinfo {author} {\bibfnamefont {K.}~\bibnamefont
  {Behnia}},\ }\href@noop {} {\emph {\bibinfo {title} {Fundamentals of
  thermoelectricity}}}\ (\bibinfo  {publisher} {OUP Oxford},\ \bibinfo {year}
  {2015})\BibitemShut {NoStop}%
\bibitem [{\citenamefont {van Houten}\ \emph {et~al.}(1992)\citenamefont {van
  Houten}, \citenamefont {Molenkamp}, \citenamefont {Beenakker},\ and\
  \citenamefont {Foxon}}]{eqat3}%
  \BibitemOpen
  \bibfield  {author} {\bibinfo {author} {\bibfnamefont {H.}~\bibnamefont {van
  Houten}}, \bibinfo {author} {\bibfnamefont {L.~W.}\ \bibnamefont
  {Molenkamp}}, \bibinfo {author} {\bibfnamefont {C.~W.~J.}\ \bibnamefont
  {Beenakker}},\ and\ \bibinfo {author} {\bibfnamefont {C.~T.}\ \bibnamefont
  {Foxon}},\ }\bibfield  {title} {\bibinfo {title} {Thermo-electric properties
  of quantum point contacts},\ }\href
  {https://doi.org/10.1088/0268-1242/7/3B/052} {\bibfield  {journal} {\bibinfo
  {journal} {Semiconductor Science and Technology}\ }\textbf {\bibinfo {volume}
  {7}},\ \bibinfo {pages} {B215} (\bibinfo {year} {1992})}\BibitemShut
  {NoStop}%
\bibitem [{\citenamefont {Mostofi}\ \emph
  {et~al.}(2008{\natexlab{b}})\citenamefont {Mostofi}, \citenamefont {Yates},
  \citenamefont {Lee}, \citenamefont {Souza}, \citenamefont {Vanderbilt},\ and\
  \citenamefont {Marzari}}]{opteq1}%
  \BibitemOpen
  \bibfield  {author} {\bibinfo {author} {\bibfnamefont {A.~A.}\ \bibnamefont
  {Mostofi}}, \bibinfo {author} {\bibfnamefont {J.~R.}\ \bibnamefont {Yates}},
  \bibinfo {author} {\bibfnamefont {Y.-S.}\ \bibnamefont {Lee}}, \bibinfo
  {author} {\bibfnamefont {I.}~\bibnamefont {Souza}}, \bibinfo {author}
  {\bibfnamefont {D.}~\bibnamefont {Vanderbilt}},\ and\ \bibinfo {author}
  {\bibfnamefont {N.}~\bibnamefont {Marzari}},\ }\bibfield  {title} {\bibinfo
  {title} {wannier90: A tool for obtaining maximally-localised wannier
  functions},\ }\href
  {https://doi.org/https://doi.org/10.1016/j.cpc.2007.11.016} {\bibfield
  {journal} {\bibinfo  {journal} {Computer Physics Communications}\ }\textbf
  {\bibinfo {volume} {178}},\ \bibinfo {pages} {685} (\bibinfo {year}
  {2008}{\natexlab{b}})}\BibitemShut {NoStop}%
\bibitem [{\citenamefont {Yates}\ \emph {et~al.}(2007)\citenamefont {Yates},
  \citenamefont {Wang}, \citenamefont {Vanderbilt},\ and\ \citenamefont
  {Souza}}]{opteq2}%
  \BibitemOpen
  \bibfield  {author} {\bibinfo {author} {\bibfnamefont {J.~R.}\ \bibnamefont
  {Yates}}, \bibinfo {author} {\bibfnamefont {X.}~\bibnamefont {Wang}},
  \bibinfo {author} {\bibfnamefont {D.}~\bibnamefont {Vanderbilt}},\ and\
  \bibinfo {author} {\bibfnamefont {I.}~\bibnamefont {Souza}},\ }\bibfield
  {title} {\bibinfo {title} {Spectral and fermi surface properties from wannier
  interpolation},\ }\href {https://doi.org/10.1103/PhysRevB.75.195121}
  {\bibfield  {journal} {\bibinfo  {journal} {Phys. Rev. B}\ }\textbf {\bibinfo
  {volume} {75}},\ \bibinfo {pages} {195121} (\bibinfo {year}
  {2007})}\BibitemShut {NoStop}%
\bibitem [{\citenamefont {Feng}\ \emph {et~al.}(2015)\citenamefont {Feng},
  \citenamefont {Guo}, \citenamefont {Zhou}, \citenamefont {Yao},\ and\
  \citenamefont {Niu}}]{mokeeq1}%
  \BibitemOpen
  \bibfield  {author} {\bibinfo {author} {\bibfnamefont {W.}~\bibnamefont
  {Feng}}, \bibinfo {author} {\bibfnamefont {G.-Y.}\ \bibnamefont {Guo}},
  \bibinfo {author} {\bibfnamefont {J.}~\bibnamefont {Zhou}}, \bibinfo {author}
  {\bibfnamefont {Y.}~\bibnamefont {Yao}},\ and\ \bibinfo {author}
  {\bibfnamefont {Q.}~\bibnamefont {Niu}},\ }\bibfield  {title} {\bibinfo
  {title} {Large magneto-optical kerr effect in noncollinear antiferromagnets
  ${\mathrm{mn}}_{3}x\phantom{\rule{0.28em}{0ex}}(x=\mathrm{Rh},\phantom{\rule{0.28em}{0ex}}\mathrm{Ir},\phantom{\rule{0.28em}{0ex}}\mathrm{Pt})$},\
  }\href {https://doi.org/10.1103/PhysRevB.92.144426} {\bibfield  {journal}
  {\bibinfo  {journal} {Phys. Rev. B}\ }\textbf {\bibinfo {volume} {92}},\
  \bibinfo {pages} {144426} (\bibinfo {year} {2015})}\BibitemShut {NoStop}%
\bibitem [{\citenamefont {Wimmer}\ \emph {et~al.}(2019)\citenamefont {Wimmer},
  \citenamefont {Mankovsky}, \citenamefont {Min\'ar}, \citenamefont {Yaresko},\
  and\ \citenamefont {Ebert}}]{mokeeq2}%
  \BibitemOpen
  \bibfield  {author} {\bibinfo {author} {\bibfnamefont {S.}~\bibnamefont
  {Wimmer}}, \bibinfo {author} {\bibfnamefont {S.}~\bibnamefont {Mankovsky}},
  \bibinfo {author} {\bibfnamefont {J.}~\bibnamefont {Min\'ar}}, \bibinfo
  {author} {\bibfnamefont {A.~N.}\ \bibnamefont {Yaresko}},\ and\ \bibinfo
  {author} {\bibfnamefont {H.}~\bibnamefont {Ebert}},\ }\bibfield  {title}
  {\bibinfo {title} {Magneto-optic and transverse-transport properties of
  noncollinear antiferromagnets},\ }\href
  {https://doi.org/10.1103/PhysRevB.100.214429} {\bibfield  {journal} {\bibinfo
   {journal} {Phys. Rev. B}\ }\textbf {\bibinfo {volume} {100}},\ \bibinfo
  {pages} {214429} (\bibinfo {year} {2019})}\BibitemShut {NoStop}%
\bibitem [{\citenamefont {Zhou}\ \emph
  {et~al.}(2019{\natexlab{a}})\citenamefont {Zhou}, \citenamefont {Hanke},
  \citenamefont {Feng}, \citenamefont {Li}, \citenamefont {Guo}, \citenamefont
  {Yao}, \citenamefont {Bl\"ugel},\ and\ \citenamefont {Mokrousov}}]{mokeeq3}%
  \BibitemOpen
  \bibfield  {author} {\bibinfo {author} {\bibfnamefont {X.}~\bibnamefont
  {Zhou}}, \bibinfo {author} {\bibfnamefont {J.-P.}\ \bibnamefont {Hanke}},
  \bibinfo {author} {\bibfnamefont {W.}~\bibnamefont {Feng}}, \bibinfo {author}
  {\bibfnamefont {F.}~\bibnamefont {Li}}, \bibinfo {author} {\bibfnamefont
  {G.-Y.}\ \bibnamefont {Guo}}, \bibinfo {author} {\bibfnamefont
  {Y.}~\bibnamefont {Yao}}, \bibinfo {author} {\bibfnamefont {S.}~\bibnamefont
  {Bl\"ugel}},\ and\ \bibinfo {author} {\bibfnamefont {Y.}~\bibnamefont
  {Mokrousov}},\ }\bibfield  {title} {\bibinfo {title} {Spin-order dependent
  anomalous hall effect and magneto-optical effect in the noncollinear
  antiferromagnets ${\mathrm{mn}}_{3}x\mathrm{N}$ with $x=\mathrm{Ga}$, zn, ag,
  or ni},\ }\href {https://doi.org/10.1103/PhysRevB.99.104428} {\bibfield
  {journal} {\bibinfo  {journal} {Phys. Rev. B}\ }\textbf {\bibinfo {volume}
  {99}},\ \bibinfo {pages} {104428} (\bibinfo {year}
  {2019}{\natexlab{a}})}\BibitemShut {NoStop}%
\bibitem [{\citenamefont {Guo}\ and\ \citenamefont {Ebert}(1995)}]{mokeeq4}%
  \BibitemOpen
  \bibfield  {author} {\bibinfo {author} {\bibfnamefont {G.~Y.}\ \bibnamefont
  {Guo}}\ and\ \bibinfo {author} {\bibfnamefont {H.}~\bibnamefont {Ebert}},\
  }\bibfield  {title} {\bibinfo {title} {Band-theoretical investigation of the
  magneto-optical kerr effect in fe and co multilayers},\ }\href
  {https://doi.org/10.1103/PhysRevB.51.12633} {\bibfield  {journal} {\bibinfo
  {journal} {Phys. Rev. B}\ }\textbf {\bibinfo {volume} {51}},\ \bibinfo
  {pages} {12633} (\bibinfo {year} {1995})}\BibitemShut {NoStop}%
\bibitem [{\citenamefont {Ravindran}\ \emph {et~al.}(1999)\citenamefont
  {Ravindran}, \citenamefont {Delin}, \citenamefont {James}, \citenamefont
  {Johansson}, \citenamefont {Wills}, \citenamefont {Ahuja},\ and\
  \citenamefont {Eriksson}}]{mokeeq5}%
  \BibitemOpen
  \bibfield  {author} {\bibinfo {author} {\bibfnamefont {P.}~\bibnamefont
  {Ravindran}}, \bibinfo {author} {\bibfnamefont {A.}~\bibnamefont {Delin}},
  \bibinfo {author} {\bibfnamefont {P.}~\bibnamefont {James}}, \bibinfo
  {author} {\bibfnamefont {B.}~\bibnamefont {Johansson}}, \bibinfo {author}
  {\bibfnamefont {J.~M.}\ \bibnamefont {Wills}}, \bibinfo {author}
  {\bibfnamefont {R.}~\bibnamefont {Ahuja}},\ and\ \bibinfo {author}
  {\bibfnamefont {O.}~\bibnamefont {Eriksson}},\ }\bibfield  {title} {\bibinfo
  {title} {Magnetic, optical, and magneto-optical properties of mnx (x=as, sb,
  or bi) from full-potential calculations},\ }\href
  {https://doi.org/10.1103/PhysRevB.59.15680} {\bibfield  {journal} {\bibinfo
  {journal} {Phys. Rev. B}\ }\textbf {\bibinfo {volume} {59}},\ \bibinfo
  {pages} {15680} (\bibinfo {year} {1999})}\BibitemShut {NoStop}%
\bibitem [{\citenamefont {Feng}\ \emph {et~al.}(2016)\citenamefont {Feng},
  \citenamefont {Guo},\ and\ \citenamefont {Yao}}]{mokeeq6}%
  \BibitemOpen
  \bibfield  {author} {\bibinfo {author} {\bibfnamefont {W.}~\bibnamefont
  {Feng}}, \bibinfo {author} {\bibfnamefont {G.-Y.}\ \bibnamefont {Guo}},\ and\
  \bibinfo {author} {\bibfnamefont {Y.}~\bibnamefont {Yao}},\ }\bibfield
  {title} {\bibinfo {title} {Tunable magneto-optical effects in hole-doped
  group-iiia metal-monochalcogenide monolayers},\ }\href
  {https://doi.org/10.1088/2053-1583/4/1/015017} {\bibfield  {journal}
  {\bibinfo  {journal} {2D Materials}\ }\textbf {\bibinfo {volume} {4}},\
  \bibinfo {pages} {015017} (\bibinfo {year} {2016})}\BibitemShut {NoStop}%
\bibitem [{\citenamefont {Zhou}\ \emph {et~al.}(2017)\citenamefont {Zhou},
  \citenamefont {Feng}, \citenamefont {Li},\ and\ \citenamefont
  {Yao}}]{mokeeq7}%
  \BibitemOpen
  \bibfield  {author} {\bibinfo {author} {\bibfnamefont {X.}~\bibnamefont
  {Zhou}}, \bibinfo {author} {\bibfnamefont {W.}~\bibnamefont {Feng}}, \bibinfo
  {author} {\bibfnamefont {F.}~\bibnamefont {Li}},\ and\ \bibinfo {author}
  {\bibfnamefont {Y.}~\bibnamefont {Yao}},\ }\bibfield  {title} {\bibinfo
  {title} {Large magneto-optical effects in hole-doped blue phosphorene and
  gray arsenene},\ }\href {https://doi.org/10.1039/C7NR05088G} {\bibfield
  {journal} {\bibinfo  {journal} {Nanoscale}\ }\textbf {\bibinfo {volume}
  {9}},\ \bibinfo {pages} {17405} (\bibinfo {year} {2017})}\BibitemShut
  {NoStop}%
\bibitem [{\citenamefont {Li}\ \emph {et~al.}(2018)\citenamefont {Li},
  \citenamefont {Zhou}, \citenamefont {Feng}, \citenamefont {Fu},\ and\
  \citenamefont {Yao}}]{mokeeq8}%
  \BibitemOpen
  \bibfield  {author} {\bibinfo {author} {\bibfnamefont {F.}~\bibnamefont
  {Li}}, \bibinfo {author} {\bibfnamefont {X.}~\bibnamefont {Zhou}}, \bibinfo
  {author} {\bibfnamefont {W.}~\bibnamefont {Feng}}, \bibinfo {author}
  {\bibfnamefont {B.}~\bibnamefont {Fu}},\ and\ \bibinfo {author}
  {\bibfnamefont {Y.}~\bibnamefont {Yao}},\ }\bibfield  {title} {\bibinfo
  {title} {Thickness-dependent magneto-optical effects in hole-doped gas and
  gase multilayers: a first-principles study},\ }\href
  {https://doi.org/10.1088/1367-2630/aabb9a} {\bibfield  {journal} {\bibinfo
  {journal} {New Journal of Physics}\ }\textbf {\bibinfo {volume} {20}},\
  \bibinfo {pages} {043048} (\bibinfo {year} {2018})}\BibitemShut {NoStop}%
\bibitem [{\citenamefont {Zhou}\ \emph
  {et~al.}(2019{\natexlab{b}})\citenamefont {Zhou}, \citenamefont {Li},
  \citenamefont {Xing},\ and\ \citenamefont {Feng}}]{mokeeq9}%
  \BibitemOpen
  \bibfield  {author} {\bibinfo {author} {\bibfnamefont {X.}~\bibnamefont
  {Zhou}}, \bibinfo {author} {\bibfnamefont {F.}~\bibnamefont {Li}}, \bibinfo
  {author} {\bibfnamefont {Y.}~\bibnamefont {Xing}},\ and\ \bibinfo {author}
  {\bibfnamefont {W.}~\bibnamefont {Feng}},\ }\bibfield  {title} {\bibinfo
  {title} {Multifield-tunable magneto-optical effects in electron- and
  hole-doped nitrogen–graphene crystals},\ }\href
  {https://doi.org/10.1039/C9TC00315K} {\bibfield  {journal} {\bibinfo
  {journal} {J. Mater. Chem. C}\ }\textbf {\bibinfo {volume} {7}},\ \bibinfo
  {pages} {3360} (\bibinfo {year} {2019}{\natexlab{b}})}\BibitemShut {NoStop}%
\end{thebibliography}%


@article{ANE,
  title = {Berry-Phase Effect in Anomalous Thermoelectric Transport},
  author = {Xiao, Di and Yao, Yugui and Fang, Zhong and Niu, Qian},
  journal = {Phys. Rev. Lett.},
  volume = {97},
  issue = {2},
  pages = {026603},
  numpages = {4},
  year = {2006},
  month = {Jul},
  publisher = {American Physical Society},
  doi = {10.1103/PhysRevLett.97.026603},
  url = {https://link.aps.org/doi/10.1103/PhysRevLett.97.026603}
}
@article{ATHE,
  title = {Energy Magnetization and the Thermal Hall Effect},
  author = {Qin, Tao and Niu, Qian and Shi, Junren},
  journal = {Phys. Rev. Lett.},
  volume = {107},
  issue = {23},
  pages = {236601},
  numpages = {5},
  year = {2011},
  month = {Nov},
  publisher = {American Physical Society},
  doi = {10.1103/PhysRevLett.107.236601},
  url = {https://link.aps.org/doi/10.1103/PhysRevLett.107.236601}
}
@article{AHE1,
  title = {Lorenz Number Determination of the Dissipationless Nature of the Anomalous Hall Effect in Itinerant Ferromagnets},
  author = {Onose, Y. and Shiomi, Y. and Tokura, Y.},
  journal = {Phys. Rev. Lett.},
  volume = {100},
  issue = {1},
  pages = {016601},
  numpages = {4},
  year = {2008},
  month = {Jan},
  publisher = {American Physical Society},
  doi = {10.1103/PhysRevLett.100.016601},
  url = {https://link.aps.org/doi/10.1103/PhysRevLett.100.016601}
}
@article{AHE2,
  title = {Extrinsic anomalous Hall effect in charge and heat transport in pure iron, ${\text{Fe}}_{0.997}{\text{Si}}_{0.003}$, and ${\text{Fe}}_{0.97}{\text{Co}}_{0.03}$},
  author = {Shiomi, Y. and Onose, Y. and Tokura, Y.},
  journal = {Phys. Rev. B},
  volume = {79},
  issue = {10},
  pages = {100404},
  numpages = {4},
  year = {2009},
  month = {Mar},
  publisher = {American Physical Society},
  doi = {10.1103/PhysRevB.79.100404},
  url = {https://link.aps.org/doi/10.1103/PhysRevB.79.100404}
}
@article{AHE3,
  title = {Effect of scattering on intrinsic anomalous Hall effect investigated by Lorenz ratio},
  author = {Shiomi, Y. and Onose, Y. and Tokura, Y.},
  journal = {Phys. Rev. B},
  volume = {81},
  issue = {5},
  pages = {054414},
  numpages = {6},
  year = {2010},
  month = {Feb},
  publisher = {American Physical Society},
  doi = {10.1103/PhysRevB.81.054414},
  url = {https://link.aps.org/doi/10.1103/PhysRevB.81.054414}
}
@article{AHE4,
  title = {The Transverse Thermomagnetic Effect in Nickel and Cobalt},
  author = {Smith, Alpheus W.},
  journal = {Phys. Rev. (Series I)},
  volume = {33},
  issue = {4},
  pages = {295--306},
  numpages = {0},
  year = {1911},
  month = {Oct},
  publisher = {American Physical Society},
  doi = {10.1103/PhysRevSeriesI.33.295},
  url = {https://link.aps.org/doi/10.1103/PhysRevSeriesI.33.295}
}
@article{AHE5,
  title = {The Hall Effect and the Nernst Effect in Magnetic Alloys},
  author = {Smith, Alpheus W.},
  journal = {Phys. Rev.},
  volume = {17},
  issue = {1},
  pages = {23--37},
  numpages = {0},
  year = {1921},
  month = {Jan},
  publisher = {American Physical Society},
  doi = {10.1103/PhysRev.17.23},
  url = {https://link.aps.org/doi/10.1103/PhysRev.17.23}
}
@article{AHE6,
  title = {Anomalous Hall Heat Current and Nernst Effect in the ${\mathrm{C}\mathrm{u}\mathrm{C}\mathrm{r}}_{2}{\mathrm{S}\mathrm{e}}_{4\ensuremath{-}x}{\mathrm{B}\mathrm{r}}_{x}$ Ferromagnet},
  author = {Lee, Wei-Li and Watauchi, S. and Miller, V. L. and Cava, R. J. and Ong, N. P.},
  journal = {Phys. Rev. Lett.},
  volume = {93},
  issue = {22},
  pages = {226601},
  numpages = {4},
  year = {2004},
  month = {Nov},
  publisher = {American Physical Society},
  doi = {10.1103/PhysRevLett.93.226601},
  url = {https://link.aps.org/doi/10.1103/PhysRevLett.93.226601}
}
@article{AHE7,
  title = {Crossover Behavior of the Anomalous Hall Effect and Anomalous Nernst Effect in Itinerant Ferromagnets},
  author = {Miyasato, T. and Abe, N. and Fujii, T. and Asamitsu, A. and Onoda, S. and Onose, Y. and Nagaosa, N. and Tokura, Y.},
  journal = {Phys. Rev. Lett.},
  volume = {99},
  issue = {8},
  pages = {086602},
  numpages = {4},
  year = {2007},
  month = {Aug},
  publisher = {American Physical Society},
  doi = {10.1103/PhysRevLett.99.086602},
  url = {https://link.aps.org/doi/10.1103/PhysRevLett.99.086602}
}
@article{AHE8,
  title = {Quantum transport theory of anomalous electric, thermoelectric, and thermal Hall effects in ferromagnets},
  author = {Onoda, Shigeki and Sugimoto, Naoyuki and Nagaosa, Naoto},
  journal = {Phys. Rev. B},
  volume = {77},
  issue = {16},
  pages = {165103},
  numpages = {20},
  year = {2008},
  month = {Apr},
  publisher = {American Physical Society},
  doi = {10.1103/PhysRevB.77.165103},
  url = {https://link.aps.org/doi/10.1103/PhysRevB.77.165103}
}

@article{yaoAHC,
  title = {Crystal Thermal Transport in Altermagnetic ${\mathrm{RuO}}_{2}$},
  author = {Zhou, Xiaodong and Feng, Wanxiang and Zhang, Run-Wu and \ifmmode \check{S}\else \v{S}\fi{}mejkal, Libor and Sinova, Jairo and Mokrousov, Yuriy and Yao, Yugui},
  journal = {Phys. Rev. Lett.},
  volume = {132},
  issue = {5},
  pages = {056701},
  numpages = {7},
  year = {2024},
  month = {Jan},
  publisher = {American Physical Society},
  doi = {10.1103/PhysRevLett.132.056701},
  url = {https://link.aps.org/doi/10.1103/PhysRevLett.132.056701}
}

@article{yaoMOKE,
  title = {Crystal chirality magneto-optical effects in collinear antiferromagnets},
  author = {Zhou, Xiaodong and Feng, Wanxiang and Yang, Xiuxian and Guo, Guang-Yu and Yao, Yugui},
  journal = {Phys. Rev. B},
  volume = {104},
  issue = {2},
  pages = {024401},
  numpages = {8},
  year = {2021},
  month = {Jul},
  publisher = {American Physical Society},
  doi = {10.1103/PhysRevB.104.024401},
  url = {https://link.aps.org/doi/10.1103/PhysRevB.104.024401}
}

@article{Altermagnetic1,
  title = {Emerging Research Landscape of Altermagnetism},
  author = {\ifmmode \check{S}\else \v{S}\fi{}mejkal, Libor and Sinova, Jairo and Jungwirth, Tomas},
  journal = {Phys. Rev. X},
  volume = {12},
  issue = {4},
  pages = {040501},
  numpages = {27},
  year = {2022},
  month = {Dec},
  publisher = {American Physical Society},
  doi = {10.1103/PhysRevX.12.040501},
  url = {https://link.aps.org/doi/10.1103/PhysRevX.12.040501}
}

@article{Altermagnetic2,
  title = {Beyond Conventional Ferromagnetism and Antiferromagnetism: A Phase with Nonrelativistic Spin and Crystal Rotation Symmetry},
  author = {\ifmmode \check{S}\else \v{S}\fi{}mejkal, Libor and Sinova, Jairo and Jungwirth, Tomas},
  journal = {Phys. Rev. X},
  volume = {12},
  issue = {3},
  pages = {031042},
  numpages = {16},
  year = {2022},
  month = {Sep},
  publisher = {American Physical Society},
  doi = {10.1103/PhysRevX.12.031042},
  url = {https://link.aps.org/doi/10.1103/PhysRevX.12.031042}
}

@article{yao3,
author = {Bai, Ling and Feng, Wanxiang and Liu, Siyuan and Šmejkal, Libor and Mokrousov, Yuriy and Yao, Yugui},
title = {Altermagnetism: Exploring New Frontiers in Magnetism and Spintronics},
journal = {Advanced Functional Materials},
volume = {34},
number = {49},
pages = {2409327},
keywords = {Altermagnetism, anomalous transport properties, nonrelativistic spin splitting, spin space group, time-reversal symmetry breaking},
doi = {https://doi.org/10.1002/adfm.202409327},
url = {https://advanced.onlinelibrary.wiley.com/doi/abs/10.1002/adfm.202409327},
abstract = {Abstract Recent developments have introduced a groundbreaking form of collinear magnetism known as “altermagnetism”. This emerging magnetic phase is characterized by robust time-reversal symmetry breaking, antiparallel magnetic order, and alternating spin-splitting band structures, yet it exhibits vanishing net magnetization constrained by symmetry. Altermagnetism uniquely integrates traits previously considered mutually exclusive to conventional collinear ferromagnetism and antiferromagnetism, thereby facilitating phenomena and functionalities previously not achievable within these traditional categories of magnetism. Initially proposed theoretically, the existence of the altermagnetic phase has since been corroborated by a range of experimental studies, which have confirmed its unique properties and potential for applications. This review explores the rapidly expanding research on altermagnets, emphasizing the novel physical phenomena they manifest, methodologies for inducing altermagnetism, and promising altermagnetic materials. The goal of this review is to furnish readers with a comprehensive overview of altermagnetism and to inspire further innovative studies on altermagnetic materials which can potentially revolutionize applications in technology and materials science.},
year = {2024}
}

@article{vasp1,
title = {Efficiency of ab-initio total energy calculations for metals and semiconductors using a plane-wave basis set},
journal = {Computational Materials Science},
volume = {6},
number = {1},
pages = {15-50},
year = {1996},
issn = {0927-0256},
doi = {https://doi.org/10.1016/0927-0256(96)00008-0},
url = {https://www.sciencedirect.com/science/article/pii/0927025696000080},
author = {G. Kresse and J. Furthmüller},
abstract = {We present a detailed description and comparison of algorithms for performing ab-initio quantum-mechanical calculations using pseudopotentials and a plane-wave basis set. We will discuss: (a) partial occupancies within the framework of the linear tetrahedron method and the finite temperature density-functional theory, (b) iterative methods for the diagonalization of the Kohn-Sham Hamiltonian and a discussion of an efficient iterative method based on the ideas of Pulay's residual minimization, which is close to an order Natoms2 scaling even for relatively large systems, (c) efficient Broyden-like and Pulay-like mixing methods for the charge density including a new special ‘preconditioning’ optimized for a plane-wave basis set, (d) conjugate gradient methods for minimizing the electronic free energy with respect to all degrees of freedom simultaneously. We have implemented these algorithms within a powerful package called VAMP (Vienna ab-initio molecular-dynamics package). The program and the techniques have been used successfully for a large number of different systems (liquid and amorphous semiconductors, liquid simple and transition metals, metallic and semi-conducting surfaces, phonons in simple metals, transition metals and semiconductors) and turned out to be very reliable.}
}

@article{vasp2,
  title = {Projector augmented-wave method},
  author = {Bl\"ochl, P. E.},
  journal = {Phys. Rev. B},
  volume = {50},
  issue = {24},
  pages = {17953--17979},
  numpages = {0},
  year = {1994},
  month = {Dec},
  publisher = {American Physical Society},
  doi = {10.1103/PhysRevB.50.17953},
  url = {https://link.aps.org/doi/10.1103/PhysRevB.50.17953}
}

@article{ggapbe,
  title = {Generalized Gradient Approximation Made Simple},
  author = {Perdew, John P. and Burke, Kieron and Ernzerhof, Matthias},
  journal = {Phys. Rev. Lett.},
  volume = {77},
  issue = {18},
  pages = {3865--3868},
  numpages = {0},
  year = {1996},
  month = {Oct},
  publisher = {American Physical Society},
  doi = {10.1103/PhysRevLett.77.3865},
  url = {https://link.aps.org/doi/10.1103/PhysRevLett.77.3865}
}

@article{wannier,
title = {wannier90: A tool for obtaining maximally-localised Wannier functions},
journal = {Computer Physics Communications},
volume = {178},
number = {9},
pages = {685-699},
year = {2008},
issn = {0010-4655},
doi = {https://doi.org/10.1016/j.cpc.2007.11.016},
url = {https://www.sciencedirect.com/science/article/pii/S0010465507004936},
author = {Arash A. Mostofi and Jonathan R. Yates and Young-Su Lee and Ivo Souza and David Vanderbilt and Nicola Marzari},
keywords = {Electronic structure, Density-functional theory, Wannier function}
}

@article{wannierberri,
  title={High performance Wannier interpolation of Berry curvature and related quantities with WannierBerri code},
  author={Tsirkin, Stepan S},
  journal={npj Computational Materials},
  volume={7},
  number={1},
  pages={33},
  year={2021},
  doi={10.1038/s41524-021-00498-5},
URL={https://doi.org/10.1038/s41524-021-00498-5},
  publisher={Nature Publishing Group UK London}
}

@article{spacegrouptable,
author = "Aroyo, Mois I. and Kirov, Asen and Capillas, Cesar and Perez-Mato, J. M. and Wondratschek, Hans",
title = "{Bilbao Crystallographic Server. II. Representations of crystallographic point groups and space groups}",
journal = "Acta Crystallographica Section A",
year = "2006",
volume = "62",
number = "2",
pages = "115--128",
month = "Mar",
doi = {10.1107/S0108767305040286},
url = {https://doi.org/10.1107/S0108767305040286},
abstract = {The Bilbao Crystallographic Server is a web site with crystallographic programs and databases freely available on-line (http://www.cryst.ehu.es). The server gives access to general information related to crystallographic symmetry groups (generators, general and special positions, maximal subgroups, Brillouin zones {\it etc.}). Apart from the simple tools for retrieving the stored data, there are programs for the analysis of group{--}subgroup relations between space groups (subgroups and supergroups, Wyckoff-position splitting schemes {\it etc.}). There are also software packages studying specific problems of solid-state physics, structural chemistry and crystallography. This article reports on the programs treating representations of point and space groups. There are tools for the construction of irreducible representations, for the study of the correlations between representations of group{--}subgroup pairs of space groups and for the decompositions of Kronecker products of representations.},
keywords = {Bilbao Crystallographic Server, point and space groups, representations of crystallographic groups},
}

@article{ruo2ahc1,
author = {Libor Šmejkal  and Rafael González-Hernández  and T. Jungwirth  and J. Sinova },
title = {Crystal time-reversal symmetry breaking and spontaneous Hall effect in collinear antiferromagnets},
journal = {Science Advances},
volume = {6},
number = {23},
pages = {eaaz8809},
year = {2020},
doi = {10.1126/sciadv.aaz8809},
URL = {https://www.science.org/doi/abs/10.1126/sciadv.aaz8809},
abstract = {Identification of a previously overlooked spontaneous Hall effect mechanism creates opportunities in low-dissipation spintronics. Electrons, commonly moving along the applied electric field, acquire in certain magnets a dissipationless transverse velocity. This spontaneous Hall effect, found more than a century ago, has been understood in terms of the time-reversal symmetry breaking by the internal spin structure of a ferromagnetic, noncolinear antiferromagnetic, or skyrmionic form. Here, we identify previously overlooked robust Hall effect mechanism arising from collinear antiferromagnetism combined with nonmagnetic atoms at noncentrosymmetric positions. We predict a large magnitude of this crystal Hall effect in a room temperature collinear antiferromagnet RuO2 and catalog, based on symmetry rules, extensive families of material candidates. We show that the crystal Hall effect is accompanied by the possibility to control its sign by the crystal chirality. We illustrate that accounting for the full magnetization density distribution instead of the simplified spin structure sheds new light on symmetry breaking phenomena in magnets and opens an alternative avenue toward low-dissipation nanoelectronics.}}

@article{ruo2ahc2,
  title={An anomalous Hall effect in altermagnetic ruthenium dioxide},
  author={Feng, Zexin and Zhou, Xiaorong and {\v{S}}mejkal, Libor and Wu, Lei and Zhu, Zengwei and Guo, Huixin and Gonz{\'a}lez-Hern{\'a}ndez, Rafael and Wang, Xiaoning and Yan, Han and Qin, Peixin and others},
  journal={Nature Electronics},
  volume={5},
  number={11},
  pages={735--743},
  year={2022},
  publisher={Nature Publishing Group UK London},
  doi = {10.1038/s41928-022-00866-z},
  URL = {https://doi.org/10.1038/s41928-022-00866-z},
}

@book{eqat1,
  title={Solid state physics},
  author={Ehrenreich, Henry and Spaepen, Frans},
  volume={56},
  year={2001},
  publisher={Academic Press}
}
@book{eqat2,
  title={Fundamentals of thermoelectricity},
  author={Behnia, Kamran},
  year={2015},
  publisher={OUP Oxford}
}
@article{eqat3,
doi = {10.1088/0268-1242/7/3B/052},
url = {https://dx.doi.org/10.1088/0268-1242/7/3B/052},
year = {1992},
month = {mar},
publisher = {},
volume = {7},
number = {3B},
pages = {B215},
author = {H van Houten and L W Molenkamp and C W J Beenakker and C T Foxon},
title = {Thermo-electric properties of quantum point contacts},
journal = {Semiconductor Science and Technology},
abstract = {The conductance, the thermal conductance, the thermopower and the Peltier coefficient of a quantum point contact all exhibit quantum size effects. The authors review and extend the theory of these effects. In addition, they review their experimental work on the quantum oscillations in the thermopower, observed using a current heating technique. New data are presented showing evidence for quantum steps in the thermal conductance, and (less unequivocally) for quantum oscillations in the Peltier coefficient. For these new experiments the authors have used a quantum point contact as a miniature thermometer.}
}

@article{opteq1,
title = {wannier90: A tool for obtaining maximally-localised Wannier functions},
journal = {Computer Physics Communications},
volume = {178},
number = {9},
pages = {685-699},
year = {2008},
issn = {0010-4655},
doi = {https://doi.org/10.1016/j.cpc.2007.11.016},
url = {https://www.sciencedirect.com/science/article/pii/S0010465507004936},
author = {Arash A. Mostofi and Jonathan R. Yates and Young-Su Lee and Ivo Souza and David Vanderbilt and Nicola Marzari},
keywords = {Electronic structure, Density-functional theory, Wannier function}
}
@article{opteq2,
  title = {Spectral and Fermi surface properties from Wannier interpolation},
  author = {Yates, Jonathan R. and Wang, Xinjie and Vanderbilt, David and Souza, Ivo},
  journal = {Phys. Rev. B},
  volume = {75},
  issue = {19},
  pages = {195121},
  numpages = {11},
  year = {2007},
  month = {May},
  publisher = {American Physical Society},
  doi = {10.1103/PhysRevB.75.195121},
  url = {https://link.aps.org/doi/10.1103/PhysRevB.75.195121}
}

@article{mokeeq1,
  title = {Large magneto-optical Kerr effect in noncollinear antiferromagnets ${\mathrm{Mn}}_{3}X\phantom{\rule{0.28em}{0ex}}(X=\mathrm{Rh},\phantom{\rule{0.28em}{0ex}}\mathrm{Ir},\phantom{\rule{0.28em}{0ex}}\mathrm{Pt})$},
  author = {Feng, Wanxiang and Guo, Guang-Yu and Zhou, Jian and Yao, Yugui and Niu, Qian},
  journal = {Phys. Rev. B},
  volume = {92},
  issue = {14},
  pages = {144426},
  numpages = {6},
  year = {2015},
  month = {Oct},
  publisher = {American Physical Society},
  doi = {10.1103/PhysRevB.92.144426},
  url = {https://link.aps.org/doi/10.1103/PhysRevB.92.144426}
}
@article{mokeeq2,
  title = {Magneto-optic and transverse-transport properties of noncollinear antiferromagnets},
  author = {Wimmer, Sebastian and Mankovsky, Sergiy and Min\'ar, J\'an and Yaresko, Alexander N. and Ebert, Hubert},
  journal = {Phys. Rev. B},
  volume = {100},
  issue = {21},
  pages = {214429},
  numpages = {11},
  year = {2019},
  month = {Dec},
  publisher = {American Physical Society},
  doi = {10.1103/PhysRevB.100.214429},
  url = {https://link.aps.org/doi/10.1103/PhysRevB.100.214429}
}
@article{mokeeq3,
  title = {Spin-order dependent anomalous Hall effect and magneto-optical effect in the noncollinear antiferromagnets ${\mathrm{Mn}}_{3}X\mathrm{N}$ with $X=\mathrm{Ga}$, Zn, Ag, or Ni},
  author = {Zhou, Xiaodong and Hanke, Jan-Philipp and Feng, Wanxiang and Li, Fei and Guo, Guang-Yu and Yao, Yugui and Bl\"ugel, Stefan and Mokrousov, Yuriy},
  journal = {Phys. Rev. B},
  volume = {99},
  issue = {10},
  pages = {104428},
  numpages = {13},
  year = {2019},
  month = {Mar},
  publisher = {American Physical Society},
  doi = {10.1103/PhysRevB.99.104428},
  url = {https://link.aps.org/doi/10.1103/PhysRevB.99.104428}
}
@article{mokeeq4,
  title = {Band-theoretical investigation of the magneto-optical Kerr effect in Fe and Co multilayers},
  author = {Guo, G. Y. and Ebert, H.},
  journal = {Phys. Rev. B},
  volume = {51},
  issue = {18},
  pages = {12633--12643},
  numpages = {0},
  year = {1995},
  month = {May},
  publisher = {American Physical Society},
  doi = {10.1103/PhysRevB.51.12633},
  url = {https://link.aps.org/doi/10.1103/PhysRevB.51.12633}
}

@article{mokeeq5,
  title = {Magnetic, optical, and magneto-optical properties of MnX (X=As, Sb, or Bi) from full-potential calculations},
  author = {Ravindran, P. and Delin, A. and James, P. and Johansson, B. and Wills, J. M. and Ahuja, R. and Eriksson, O.},
  journal = {Phys. Rev. B},
  volume = {59},
  issue = {24},
  pages = {15680--15693},
  numpages = {0},
  year = {1999},
  month = {Jun},
  publisher = {American Physical Society},
  doi = {10.1103/PhysRevB.59.15680},
  url = {https://link.aps.org/doi/10.1103/PhysRevB.59.15680}
}

@article{mokeeq6,
doi = {10.1088/2053-1583/4/1/015017},
url = {https://dx.doi.org/10.1088/2053-1583/4/1/015017},
year = {2016},
month = {nov},
publisher = {IOP Publishing},
volume = {4},
number = {1},
pages = {015017},
author = {Feng, Wanxiang and Guo, Guang-Yu and Yao, Yugui},
title = {Tunable magneto-optical effects in hole-doped group-IIIA metal-monochalcogenide monolayers},
journal = {2D Materials}
}

@Article{mokeeq7,
author ="Zhou, Xiaodong and Feng, Wanxiang and Li, Fei and Yao, Yugui",
title  ="Large magneto-optical effects in hole-doped blue phosphorene and gray arsenene",
journal  ="Nanoscale",
year  ="2017",
volume  ="9",
issue  ="44",
pages  ="17405-17414",
publisher  ="The Royal Society of Chemistry",
doi  ="10.1039/C7NR05088G",
url  ="http://dx.doi.org/10.1039/C7NR05088G",}

@article{mokeeq8,
doi = {10.1088/1367-2630/aabb9a},
url = {https://dx.doi.org/10.1088/1367-2630/aabb9a},
year = {2018},
month = {apr},
publisher = {IOP Publishing},
volume = {20},
number = {4},
pages = {043048},
author = {Li, Fei and Zhou, Xiaodong and Feng, Wanxiang and Fu, Botao and Yao, Yugui},
title = {Thickness-dependent magneto-optical effects in hole-doped GaS and GaSe multilayers: a first-principles study},
journal = {New Journal of Physics}}
@Article{mokeeq9,
author ="Zhou, Xiaodong and Li, Fei and Xing, Yanxia and Feng, Wanxiang",
title  ="Multifield-tunable magneto-optical effects in electron- and hole-doped nitrogen–graphene crystals",
journal  ="J. Mater. Chem. C",
year  ="2019",
volume  ="7",
issue  ="11",
pages  ="3360-3368",
publisher  ="The Royal Society of Chemistry",
doi  ="10.1039/C9TC00315K",
url  ="http://dx.doi.org/10.1039/C9TC00315K"}
%

\end{document}